\documentclass[nofootinbib]{article}
\usepackage[total={6.5in,9in},top=1in,headsep=0.1in,headheight=1in]{geometry}
\usepackage[utf8]{inputenc}
\usepackage{hyperref}
\usepackage{amssymb}
\usepackage{graphicx}
\usepackage{lipsum}
\usepackage{hyperref}
\usepackage{placeins}
\usepackage[maxbibnames=10,style=numeric-comp,sorting=none,backend=bibtex]{biblatex}
\usepackage{upgreek}
\hypersetup{hidelinks}
\usepackage[colorinlistoftodos]{todonotes}
\usepackage[hybrid]{markdown}
\usepackage{color}
\usepackage{enumitem}
\usepackage[caption=false]{subfig}

\newcommand\snowmass{
\begin{center}
  \rule[-0.2in]{\hsize}{0.01in}\\
  \rule{\hsize}{0.01in}\\
  \vskip 0.1in
  Submitted to the Proceedings of the US Community Study\\ 
  on the Future of Particle Physics (Snowmass 2021)\\
  \rule{\hsize}{0.01in}\\
  \rule[+0.2in]{\hsize}{0.01in}\\[-2em]
\end{center}
}

\usepackage[firstpage=true]{background}
\backgroundsetup{contents={\parbox{6.5in}{\snowmass}}, scale=1,placement=top,opacity=1,color=black,position={3.25in,1.2in}}

\usepackage{fancyhdr}
\fancypagestyle{plain}{%
  \fancyhf{}%
  \fancyhead[C]{}
  \fancyfoot[C]{\thepage}
}

\fancypagestyle{empty}{%
  \fancyhf{}%
  \fancyhead[C]{{\it Snowmass 2021 Cross Frontier Report: Dark Matter Complementarity (Extended Version)}}
  \fancyfoot[C]{\thepage}
}
\title{Snowmass 2021 Cross Frontier Report: Dark Matter Complementarity (Extended Version)}
\date{\today}

\usepackage{authblk}

\author[1]{Antonio Boveia}

\author[2]{Mohamed Berkat}

\author[3]{Thomas Y. Chen}

\author[ \hspace{-1ex}]{Aman Desai}

\author[2,4]{Caterina Doglioni}

\author[5,6,7]{Alex Drlica-Wagner}

\author[8]{Susan Gardner}

\author[9]{Stefania Gori}

\author[2]{Joshua Greaves}

\author[10]{Patrick Harding}

\author[11]{Philip C. Harris}

\author[12]{W. Hugh Lippincott}

\author[13,14,15]{Maria Elena Monzani}

\author[16]{Katherine Pachal}

\author[17]{Chanda Prescod-Weinstein}

\author[18]{Gray Rybka}

\author[19]{Bibhushan Shakya}

\author[20]{Jessie Shelton}

\author[21]{Tracy R. Slatyer}

\author[22]{Amanda Steinhebel}

\author[23]{Philip Tanedo}

\author[13]{Natalia Toro}

\author[13]{Yun-Tse Tsai}

\author[11]{Mike Williams}

\author[11]{Lindley Winslow}

\author[24]{Jaehoon Yu}

\author[25]{Tien-Tien Yu}

\affil[1]{Department of Physics and Center for Cosmology and Astroparticle Physics, The Ohio State University,
191 W. Woodruff Avenue Columbus, OH 43210, USA}

\affil[2]{Fysiska institutionen, Lunds universitet, Professorsgatan 1, Lund, Sweden}

\affil[3]{Fu Foundation School of Engineering and Applied Science, Columbia University, New York, NY 10027, USA}

\affil[4]{University of Manchester, Department of Physics and Astronomy, Manchester M13 9PL, United Kingdom}

\affil[5]{Fermi National Accelerator Laboratory, Batavia, IL 60510, USA}
\affil[6]{Kavli Institute for Cosmological Physics, University of Chicago, Chicago, IL 60637, USA}
\affil[7]{Department of Astronomy and Astrophysics, University of Chicago, Chicago IL 60637, USA}

\affil[8]{Department of Physics and Astronomy, University of Kentucky, Lexington, KY 40506-0055}

\affil[9]{Physics Department, University of California, Santa Cruz, CA 95064, USA}

\affil[10]{Physics Division, Los Alamos National Laboratory, Los Alamos, NM 87545, USA}

\affil[11]{Laboratory for Nuclear Science, Massachusetts Institute of Technology, Cambridge, MA 02139, USA}

\affil[12]{Department of Physics, University of California, Santa Barbara, CA 93106, USA}

\affil[13]{SLAC National Accelerator Laboratory, Menlo Park, CA 94025, USA}

\affil[14]{Kavli Institute for Particle Astrophysics and Cosmology, Stanford University, Stanford CA, USA}

\affil[15]{Vatican Observatory, Castel Gandolfo, V-00120, Vatican City State}

\affil[16]{TRIUMF, Vancouver, BC V6T 2A3, Canada}

\affil[17]{Department of Physics and Astronomy, University of New Hampshire, Durham, NH 03824, USA}

\affil[18]{Department of Physics, University of Washington, Seattle WA 98195, USA}

\affil[19]{Deutsches Elektronen-Synchrotron DESY, Notkestr.~85, 22607 Hamburg, Germany}

\affil[20]{Department of Physics, University of Illinois Urbana-Champaign, Urbana, IL 61801}

\affil[21]{Center for Theoretical Physics, Massachusetts Institute of Technology, Cambridge, MA 02139, USA}

\affil[22]{NASA Goddard Space Flight Center, Greenbelt, MD, USA}

\affil[23]{Department of Physics and Astronomy, University of California Riverside, Riverside, CA 92521, USA}

\affil[24]{Physics Department, University of Texas, Arlington, TX 76019, USA}

\affil[25]{Department of Physics and Institute for Fundamental Science, University of Oregon, Eugene, OR 97403, USA}

\begin{document}
\maketitle

\begin{abstract}
The fundamental nature of Dark Matter is a central theme of the Snowmass 2021 process, extending across all frontiers. In the last decade, advances in detector technology, analysis techniques and theoretical modeling have enabled a new generation of experiments and searches while broadening the types of candidates we can pursue. Over the next decade, there is great potential for discoveries that would transform our understanding of dark matter. In the following, we outline a road map for discovery developed in collaboration among the frontiers. A strong portfolio of experiments that delves deep, searches wide, and harnesses the complementarity between techniques is key to tackling this complicated problem, requiring expertise, results, and planning from all Frontiers of the Snowmass 2021 process. 

\end{abstract}

\newpage

\section*{Executive Summary}
The evidence for Dark Matter (DM) is overwhelming, yet the fundamental nature of its constituents remains a mystery. Over the last decade, we have built a powerful and diverse collection of tools to unlock this mystery, both by refining established technologies and techniques and by harnessing new ones including artificial intelligence/machine learning (AI/ML) and quantum sensing/control. At the same time, we have continued to build our understanding of how DM shapes our universe. We are well-positioned for a great discovery.

From its production to its interactions, DM is a major science driver across all experimental frontiers: Cosmic Frontier (CF), Energy Frontier (EF), Rare and Precision Frontier (RF) and Neutrino Frontier (NF), as well as the cross-cutting frontiers Accelerator Frontier (AF), Community Engagement Frontier (CEF), Computing Frontier (CompF), Instrumentation Frontier (IF), Underground Facilities (UF) and Theory Frontier (TF). Because the science of DM does not respect frontier boundaries, a unified strategy is needed to maximize discovery potential. This is done by understanding the complementarity between techniques and technologies in order to enable discoveries. Furthermore, this complementarity is necessary to characterize the nature of any putative DM candidate that is discovered.

\noindent
{\bf Complementarity Within and Across Frontiers}

Complementarity drives discovery in multiple ways. The space of viable DM candidates and their properties is very large, and consequently a single experimental technique or approach cannot be used to test all the possibilities; a diverse range of techniques provides access to a much broader ensemble of DM scenarios. Furthermore, where different approaches have simultaneous sensitivity to a particular DM candidate, they will provide essential and complementary information and promote healthy competition. Different techniques will shed light on different properties of DM -- such as telling us whether a new particle constitutes the bulk of the DM in the Galactic halo, elucidating the interactions of DM with known particles, and potentially mapping the spectrum of other new ``dark sector'' particles related to the DM. In the event of a discovery, detection and exclusion by complementary techniques will help triangulate the fundamental nature of DM.



\noindent
{\bf Maximize Opportunities for Discovery: Delve Deep, Search Wide}

Embracing the role of complementarity, the DM community proposes a strategy to \emph{delve deep and search wide} to maximize discovery potential. A range of highly compelling theoretical targets arising from simple/minimal models are accessible in the next decade via planned and proposed CF, EF, NF, and RF experiments, colliders, and observatories. 
While discovering the fundamental nature of DM is the ultimate prize, searching in these regions and \emph{not} finding DM would provide important information on the properties of DM. Simultaneously, our strategy encompasses the development of new technologies and techniques to explore an even broader field of possibilities and complement the sensitivity of existing searches.

\noindent
{\bf Discovery Strategy}\\
The community puts forth the following strategy for discovering the fundamental nature of DM:
\begin{itemize}
\item \textbf{Build a portfolio of experiments of different scales:}
Experiments at all scales are needed to untangle the mystery of DM and cover the very broad range of theoretically motivated parameter space. Existing and planned large-scale facilities across the HEP frontiers have exceptional potential to discover the fundamental properties of DM.
We should commit to scaling up mature technologies that can promise significant sensitivity improvements, developing potentially transformative new technologies to maturity, and supporting efforts to maximize and make accessible large projects' science output in the search for DM.
At smaller scales, execution of the existing Dark Matter New Initiatives program and similar future calls are necessary to build the most compelling DM portfolio, develop experience in project execution, and accelerate the pace of discovery. 

\item \textbf{Leverage US expertise in international projects:}
The effort to understand the fundamental nature of DM is a world-wide endeavor. Coordination and cooperation across borders is critical for enabling this discovery. While building a strong US-based program, we should pursue opportunities to leverage key US expertise as a collaborative partner in international projects and play a leadership role in this critical area.  

\item \textbf{Provide support to further strengthen the theory program:}
A strong theoretical program is essential to make connections between experimental frontiers and leverage new developments in analysis techniques. Theorists' input has been and will be critical for developing innovative new approaches to better understand and detect DM, as well as determining how to predict and relate signals across a range of experimental probes.

\item \textbf{Support inter-disciplinary collaborations that enable discovery:}
Many searches for DM benefit greatly from cross-disciplinary expertise with examples ranging from nuclear physics to metrology, and astrophysics to condensed matter and atomic physics. Mechanisms to support such inter-disciplinary collaborations should be established.


\item\textbf{Targeted increase in the research budget:} 
New research funding targeted toward solving the DM problem is essential to enable new ideas, new technologies, and new analyses. The number of active efforts exploring DM has increased tremendously in the past decade, without a concomitant increase in research funding. 
Across all frontiers and project scales, research funding is critical to enable discovery and leverage new capabilities, both in projects focused specifically on DM and to support DM analyses at multi-purpose experiments.
Without such support, the community will not be able to execute the program described here, decreasing the chances of solving the mystery of DM. 





\end{itemize}

\newpage
\section{Introduction}

Determining the fundamental nature of dark matter (DM) is one of the major open questions that confronts our understanding of physics, and one that has guided much of the Snowmass process in most HEP Frontiers. 
While the microscopic properties of DM remain almost completely unknown, the relatively similar energy densities of dark and visible matter in the Universe --- DM has five times more energy density than visible matter --- can be taken as compelling evidence of the existence of non-gravitational interactions between DM and the Standard Model (SM). However, the nature of these interactions is essentially unconstrained, necessitating a broad and comprehensive approach to this question to make progress in the next decade. 
This requires expertise, results, and planning from the theoretical, computational, experimental, instrumentation, and accelerator communities and involves nearly all Frontiers of the Snowmass 2021 process. Taking advantage of synergies between different DM search strategies is also recommended by the European Particle Physics Strategy Update \cite{Strategy:2019vxc,European:2720131}. 

The 2013 Snowmass process had a topical group (CF4) specifically devoted to the complementarity of different DM studies. 
The white paper produced by that group, \textit{``Dark Matter in the Coming Decade: Complementary Paths to Discovery and Beyond,"}\cite{Snowmass2013CosmicFrontierWorkingGroups1-4:2013wfj}
reviewed existing and planned DM efforts in direct detection, indirect detection and collider experiments, as well as in astrophysical probes. 
Using two simple theoretical frameworks for quantitative comparisons, this white paper highlighted the complementarity of these different DM search programs.  

The need for diverse and complementary approaches to the DM problem is even more pressing now than it was in 2013.
The DM search domain has broadened significantly; many promising new avenues for understanding DM are being developed now and will yield results in the next decade.
Theoretical understanding of possible DM parameter space has grown, and it is being explored by a much larger number of projects at different scales. Working in tandem, experiment, observation, theory, and computation have the potential to identify key dark matter properties while definitively excluding vast swathes of parameter space.


While the stated need for complementarity presented in 2013 still stands, this document 
re-casts the scope and definition of complementary approaches to DM identification in terms of the current state of the field (Section \ref{sec:DMComplementarity}), 
 summarizes the needs of the different communities looking for DM and their complementary strengths (Section \ref{sec:IndividualTGComplementarityNeeds}),
and supports  these arguments with cross-Frontier case studies (Section \ref{sec:CaseStudies}). 

\section{Dark matter complementarity in the coming decade}
\label{sec:DMComplementarity}

Different approaches to DM searches are necessary and complementary for the following reasons: 

\begin{enumerate}

\item \textbf{Different experiments can simultaneously discover the properties of dark matter by detecting relic dark matter and by producing it in the lab}. 
Relic particle or wave-like DM  that is already present in the universe can be detected by direct, indirect, and cosmic searches in the Cosmic Frontier. DM production could be observed in extreme environments in the Cosmic Frontier, or under controlled laboratory conditions at accelerators and colliders in the Energy, Rare Processes and Precision, and Neutrino Frontiers. Probes that detect relic DM give insight into the properties of the DM halo and provide the connection of new particle discoveries with cosmological DM, while probes studying DM production give insight on the DM interactions including its early-universe behaviour and on the dark sector particle spectrum beyond DM. 
This kind of complementarity is showcased in the case studies \textit{Minimal WIMP} (Section \ref{sub:minimalWIMP}), involving the Energy and Cosmic Frontiers, \textit{Sterile Neutrino} (Section \ref{sub:SterileNeutrino}) involving the Cosmic and Neutrino Frontiers, and \textit{Axion DM} (Section \ref{sub:axion}), involving the Cosmic, Neutrino, Rare Processes, and Precision Frontiers. 

\item \textbf{Different approaches to search for DM have unique discovery sensitivity to specific regions of parameter space.} It is by performing a wide variety of experiments from different Frontiers that we can span the wide DM parameter space. This kind of complementarity is exemplified in \textit{Case Study: Vector Portal} (Section \ref{sub:BSMMediation}), involving the Energy and Rare Processes and Precision Frontier. 
It is also found in the complementarity between DM detection experiments sensitive to non-relativistic DM signals (at low velocity/low energy) and production probes that explore the physics of DM at higher energy and in the relativistic regime.

\item \textbf{Different probes have different strengths and offer sensitivity to different DM properties}. 
For example, terrestrial probes enjoy controllable and clean environments, and typically are sensitive to rare DM interaction processes; cosmological and astrophysical probes allow access to environments not found on Earth and time/space scales dwarfing terrestrial experiments, and consequently have unique sensitivity to properties like lifetime, DM self-interaction cross section, etc. 
This kind of complementarity is shown in \textit{Case Study: Sterile Neutrino DM} (Section \ref{sub:SterileNeutrino}), involving the Cosmic and Energy Frontiers and the Neutrino Frontier. 

\item \textbf{Different DM experiments can be co-located and/or profit from the same or similar technological infrastructure.} In practice, this means a wider exploration of DM with a more efficient use of shared resources. This is the case of smaller Rare Processes and Precision Frontier accelerator experiments that can be co-located with Energy Frontier collider experiments, using the same high-energy beams to produce different kinds of DM. Such experiments are discussed in depth in the Rare and Precision Frontier and their connection to collider experiments is discussed in the Energy Frontier report~\cite{Energy-Frontier-Report}.  

\end{enumerate}

\textbf{The community searching for DM has grown much more diversified in terms of technologies, search targets, and project scales.} 
Since the last iteration of Snowmass in 2013, many new approaches to searching for DM, as well as new theoretical hypotheses, have attained sufficient maturity to be part of the toolkit that we will use to make progress in the quest for DM in the next decade. 
The older approach to DM complementarity (as outlined in the previous Snowmass whitepaper \cite{Snowmass2013CosmicFrontierWorkingGroups1-4:2013wfj}) focused primarily on the WIMP hypothesis and direct DM-SM interactions, and incorporated high-mass new particle searches, direct and indirect detection, and astrophysical probes. This remains an important hypothesis that should be rigorously tested as part of a program that ``delves deep.'' However, the WIMP is now joined by a greater diversity of alternative DM candidates. The QCD axion has emerged as another key target for a focused suite of experiments that will enable a definitive search for this candidate. Candidates and search strategies that form the wider strategy include:




\begin{itemize}[leftmargin=1.0em, noitemsep]
\item Light particle-like DM with masses in the MeV-GeV range,
\item Wave-like DM beyond the QCD axion such as ultralight axion-like particles,
\item Signatures of the greater dark sector including long-lived particles that use new and existing detectors and accelerators,
\item Cosmological observations of DM's interaction on large scales,
\item New signatures of DM from gravitational wave detection and other multi-messenger sources. 
\end{itemize}

The strategy presented above was developed from the bottom-up through the communities represented by the Frontier Topical Groups. It is well-aligned with the Basic Research Needs for Dark Matter Small Projects New Initiatives report \cite{BRNreport}, which highlighted the growing landscape of smaller experiments to produce and detect DM at accelerators, as well as the direct detection of light and ultra-light DM. The goal of the following sections is to summarize the needs of the individual Topical Groups and show that the needs are highly complementary as are the techniques. A common strategy across HEP is needed to enable the discoveries that will reveal the nature of DM.

\section{Realizing dark matter complementarity across Frontiers}
\label{sec:IndividualTGComplementarityNeeds}

In this section, we briefly summarize the main approaches towards identifying the particle nature of DM from each Snowmass Frontier and their needs, referring to the Topical Group whitepapers for further information. We also outline their unique strengths and capabilities as well as their synergies with approaches in other Frontiers. 


\subsection{Cosmic Frontier 1 - Direct and Indirect Searches for Particle Dark Matter}

There are many theoretically motivated DM candidates capable of explaining the astrophysical evidence for DM, which span mass ranges from the lightest masses that are still consistent with galactic structure ($m_\chi\sim 10^{-21}$ eV) to macroscopic objects that can be several hundreds of solar masses. The Cosmic Frontier 1 topical group focused on DM candidates with masses from the eV range to the Planck scale ($m_\chi\sim M_{\rm Pl}=10^{19}$ GeV).
In this mass range, DM candidates are best described by a particle picture and we denote these candidates as {\bf particle DM}. Understanding the {\it particle} nature of such DM candidates will require multiple experimental techniques spanning the entire range of the High-Energy Physics (HEP) program, including both direct and indirect detection techniques. 


The P5 recommendations of 2014 included increased funding of the then-proposed generation of direct detection experiments (sometimes called ``Generation 2
" or ``G2''), support for one or more next generation experiments, maintaining a program at smaller scale to support new ideas and developments, and possible investment in the Cherenkov Telescope Array (CTA) experiment. The G2 experiments are now in operation or near the end of construction, but unfortunately there has been no movement in the US towards a next generation program. In 2018, the Basic Research Needs Dark Matter New Initiatives program (DMNI) ~\cite{BRNreport} led to the development of several promising direct detection ideas; this process was very valuable to the community and provides a useful model for supporting smaller projects of all kinds going forward. Unfortunately, the current DMNI experiments also await the next stage of funding to proceed to the project phase.  

\vspace{1em}
\noindent \textbf{Scientific Opportunities \& Roadmap} 

The next decade offers a broad array of exciting opportunities in direct and indirect detection, as discussed in Ref.~\cite{Cooley:2022ufh}.

On that timescale, future-generation direct detection experiments will have the capacity to probe spin-independent nuclear scattering of heavy DM candidates in the parameter space that reaches all the way to the neutrino ``fog'', the expected background from the coherent elastic neutrino-nucleus scattering of solar and atmospheric neutrinos. Simultaneously, other searches will advance sensitivity by an order of magnitude beyond the reach of current generation experiments in scenarios where the fog remains many orders of magnitude distant, such as spin-dependent interactions. This improved sensitivity will test a range of highly-motivated WIMP models that remain currently unexplored (one set of examples is discussed in Section~\ref{sub:minimalWIMP}). 

At lower masses, a compelling case for light DM particles has been made from a theoretical standpoint over the past decade,
and new techniques have been developed to match. A new wave of technological development has enabled direct-detection experiments with sensitivity to tiny energy depositions, at the eV scale and below. This progress will allow current and upcoming experiments to probe scattering of DM as light as 1 MeV and absorption of DM as light as 1 eV.

Indirect detection has near-term potential to provide model-agnostic probes of the minimal thermal relic scenario with $s$-wave annihilation up to tens of TeV masses, delving deep into WIMP parameter space. Upcoming and proposed searches would also provide the first dedicated probe of low-energy cosmic-ray antideuterons as a possible new low-background discovery channel, address a current gap in sensitivity in the MeV gamma-ray band, provide new tests of sterile neutrino DM, and more generally enhance our sensitivity to DM across a very broad range of energy scales and cosmic messengers.

To pursue these opportunities and maximize the probability of a transformative discovery, the CF01 community has advanced the following principles:

\begin{itemize} 
\item A diverse, continuous portfolio of experiments that includes both direct and indirect detection techniques at multiple scales maximizes the probability of discovering particle DM.
\item Moderate- and large-scale experiments allow us to delve deep into high priority target scenarios such as WIMPs, whereas a well-chosen ensemble of small-scale experiments provides versatility and the ability to test an expanded range of models. 
\item For direct detection in particular, it is key to support both a mid-to-large scale next-generation program and a portfolio of smaller scale experiments as in the DMNI program.
\item Support for theory, simulations, calibration, background modeling and complementary astrophysical measurements as essential to enabling discovery. 
\item R\&D towards improved detector technologies are fundamental and crucial to a comprehensive exploration of the dark sector.
\item Direct detection experiments, particularly the next generation of WIMP searches, will require continued investment in underground facilities.
\end{itemize}

\vspace{1em}
\noindent {\bf Particular Strengths and Capabilities}

 Direct detection experiments using mature technologies like large liquid noble detectors provide ultra-clean, low-temperature, controlled environments that allow us to probe some of the tiniest cross sections ever measured. Experimentalists have succeeded in creating the ``cleanest environment in the known universe.'' These experiments probe DM in the non-relativistic regime -- favorable for models where interactions are enhanced at low velocities, and providing leading sensitivity to heavy DM up to ultraheavy mass scales. Direct detection experiments are adaptable, and can respond readily to signals or hints or the lack thereof to mitigate systematic backgrounds or redirect to new parameter space. Furthermore, direct detection experiments can be, and often are, adjusted, upgraded, fixed, or moved to confirm or disprove any observation of DM that may arise.

Indirect detection searches offer sensitivity to DM from the keV scale to the Planck scale, the definition of ``searching wide''. As the volumes and timescales accessible to indirect searches dwarf those of any terrestrial detector, such searches also provide unique probes of long-timescale processes such as DM decay (and other such dark-sector processes, e.g. involving metastable/weakly-coupled mediators) over enormous mass ranges. Especially for higher DM masses, indirect signals are generically expected to be multi-messenger and multi-scale and multiwavelength, allowing powerful internal consistency checks on a putative signal. In broad classes of models, most notably classic thermal relics, indirect signals are directly tied to the mechanism fixing the DM abundance and thus provide broadly model-agnostic tests of that mechanism.  

Perhaps most importantly, cosmic frontier particle searches probe for DM within its astrophysical environment, connecting any observed candidate to the cosmological evidence that motivates the entire DM program in the first place. For example, detection would allow powerful probes of DM halo properties. Any detection that arises through the cosmic frontier can provide a target and motivation for future efforts in the other frontiers.

\vspace{1em}
\noindent {\bf Complementarity}

Particle searches in the cosmic frontier have overlaps with almost every other frontier in particle physics. One clear connection is to the instrumentation frontier. Developments in instrumentation are what drive increases in sensitivity. Two obvious areas are efforts to improve liquid noble detectors, which has driven huge sensitivity increases in searching for the WIMP and promise to allow searches all the way to the neutrino fog in the next generation. Developments in quantum sensing, CCDs, and phonon detection are leading to an explosion of searches for lighter DM, with sensitivity that simply wasn't imaginable ten years ago. 

Direct detection experiments take place in deep underground laboratories, requiring a strong connection to Underground Facilities. In the next decade, these facilities will need to continue to expand to make room for larger next generation liquid noble detectors, while also increase their cryogenic capabilities to enable the new wave of technologies to run in low background environments. 

In terms of sensitivity, CF particle searches share information with all the DM searches going on in the energy, neutrino, and rare processes frontiers. Indirect detection has sensitivity to neutrino interactions, and boosted DM searches offer sensitivity to complementary parameter space. Accelerator experiments like LDMX are complementary to low-threshold direct detection experiments; the former offers significant reach to light DM models with MeV-scale mediators while the latter excels for models in which the interaction cross-sections are enhanced at low momentum. Accelerator experiments are also complementary to indirect searches for light DM, again because they allow probes of number-changing interactions that are suppressed in the non-relativistic regime. Combining the results of these efforts provides the most information about any potential DM candidate. Energy frontier experiments can similarly measure properties of DM that no cosmic probe could resolve, and also provide a complementary sensitivity. As is the theme of this report, all the frontiers offer different ways of seeing DM, and only by putting them all together can we have  a complete picture. 

\subsection{Cosmic Frontier 2 - Wavelike-Dark Matter: The QCD Axion and Beyond}

Wave-like DM (WLDM) encompasses all candidates with masses less than 1\,eV. Due to their small masses, the detection principles are vastly different than those traditionally used in high energy physics. It is here where quantum measurement techniques become critical and advancements in this area have opened up a broad horizon of new candidates to explore and many opportunities for discovery. 

Within this group is the well-motivated QCD axion. This is where the community proposes to delve deep. Originally proposed to solve the strong CP problem, it is also an excellent DM candidate with both pre- and post-inflation production scenarios. The pseudo-scalar QCD axion exists in a specific class of models, but a broader class of theories with higher dimension operators often produce similar Axion-Like-Particles (ALPs). The first detection in an axion experiment would start a race to determine its couplings to the SM.   

As part of the 2014 P5, the flagship axion experiment ADMX was funded as part of the G2 suite of experiments. Since then, the community has grown exponentially motivated by advances in both key technologies and theory with many demonstrator-scale experiments producing exciting results. The DMNI program has been particularly critical to this success and allowed two experiments DMRadio-m$^3$ and ADMX-EFR to finalize designs and develop project execution plans. Continued support for the DMNI program and small-to-mid scale experiments is critical for enabling discovery in this area.

\vspace{1em}
\noindent \textbf{Scientific Opportunities \& Roadmap} 

This growing community's needs are modest, but care will be needed to ensure that the US capitalizes on its investment and leadership in this area. The community has come together behind a roadmap that is in line with the strategy of delving deep and searching wise. The key points are:

\begin{itemize}[leftmargin=1.0em]
\item {\bf Pursue the QCD Axion by Executing the Current Projects}
The ADMX G2 effort continues to scan exciting axion DM parameter space and the experiments DMRadio-m3 and ADMX-EFR are prepared to start executing their project plans.

\item {\bf Pursue WLDM with a Collection of Small-Scale Experiments}
The search for WLDM requires a variety of techniques. The community would benefit from a concerted effort to foster small scale projects. The DOE DMNI process has worked very well for this.  In addition, we should pursue opportunities to harness key US expertise for International projects.

\item {\bf Support Enabling Technologies and Cross Disciplinary Collaborations}
Common needs include ultra-sensitive quantum measurement and quantum control, large high-field magnets, spin ensembles, and sophisticated resonant systems. These have strong synergies with other HEP needs.

\item {\bf Support Theory Beyond the QCD Axion}
The QCD axion is an important benchmark model, but not the only motivated one. Theoretical effort should be supported to understand the role of scalars, vectors and ALPs in DM cosmology and astrophysics and to explore new detection modalities.
\end{itemize}

\vspace{1em}
\noindent
{\bf Particular Strengths and Capabilities}


As with particle-like DM, direct detection of WLDM would provide immediate information about our astrophysical environment. In most WLDM detection scenarios, the detection can immediately be followed up with a precision measurement of their velocity distribution in the halo and the directional signal that can be used to further inform our understanding of the halo's structure. This will allow refined models of galaxy formation and have consequences for our understanding of structure evolution in our universe.

Unlike most particle-like DM, the production mechanisms of WLDM typically occur around the time of inflation. Thus information (in particular the mass scale) about WLDM may be used to constrain the energy scale of inflation. In this sense, despite the small masses involved, WLDM detection indirectly probes energy scales much higher than would be otherwise accessible. Conversely, signatures in inflation (such as the presence or absence of B-modes in the cosmic microwave background) may be used to constrain most-likely mass ranges of WLDM for detectors to target.

The detection technology of WLDM is well-aligned with national and global efforts to improve quantum technologies. The detection of WLDM is inherently quantum, and this generation of detectors pioneered the use of quantum techniques for HEP (e.g. quantum-limited SQUIDs and JPA in ADMX). The envisioned next generation of experiments will push further, exploring squeezing and counting technologies vital for quantum computer development. WLDM detectors also demand advances in the fields of large-scale cryogenics and high-field magnet technology, with significant overlap with the needs of next generation accelerators.


\vspace{1em}
\noindent
{\bf Complementarity}

For wave DM to date, complementarity has mostly focused on how astrophysical and cosmic probes constrain the available mass range particularly for the QCD axion. However, the phenomenology of WLDM is rich, and its detection would be proof of the existence of additional heavy particles and higher order theories. The above recommendation for more theory support is to better understand the connections to the Rare Processes, Neutrino and Energy Frontiers, and how a WLDM discovery would guide both precision measurements and future colliders. There are also profound connections between these fields existing in the early universe and cosmological observables, especially for the CMB B-modes.

The complementarity within instrumentation is more immediate. The techniques needed for discovery have strong overlap with RF table top experiments such as fifth force experiments and searches for EDMs. Both quantum and traditional technologies not only have a strong overlap with other Cosmic Frontier efforts but also with magnet, cryogenic and RF cavity technology found on the Accelerator Frontier.  

\subsection{Cosmic Frontier 3 - Cosmic Probes of Dark Matter}

Cosmic probes provide the only direct, positive empirical measurements of DM's existence and properties.
These probes complement terrestrial DM searches by constraining the interaction strength between DM and the SM in otherwise inaccessible regions of parameter space.
In addition, cosmic probes provide the only known way to directly study the fundamental properties of DM through gravity, the only force to which DM is known to couple.
Cosmic probes are sensitive to the DM mass, lifetime, self-interaction cross section, and other dark sector particles.


Cosmic probes have emerged as an important avenue to measure the fundamental properties of DM. 
Current/near-future HEP cosmology experiments have direct sensitivity to DM particle physics. 
Cosmological studies of DM have a unique ability to probe DM microphysics and link the results of terrestrial DM experiments to cosmological measurements.
The construction of future cosmology experiments is critical for expanding our understanding of DM physics. Proposed facilities across the electromagnetic spectrum, as well as gravitational waves, can provide sensitivity to DM physics, as well as physics of dark energy and the early universe. 
Strategic HEP investments in the construction and operation of these facilities can optimize their sensitivity to DM physics as a core consideration in their design.
Cosmic probes provide robust sensitivity to the microphysical properties of DM due to enormous progress in theoretical modeling, numerical simulations, and astrophysical observations. 
Theory, simulation, observation, and experiment must be supported together to maximize the efficacy of cosmic probes of DM physics.


\vspace{1em}
\noindent \textbf{Scientific Opportunities \& Roadmap}

\noindent Cosmic probes have emerged as a new field in the endeavor to measure the fundamental, microscopic properties of DM. Three core priorities leverage and complement historical HEP community strengths and capabilities.

\begin{itemize}[nosep]
    \item Current/near-future HEP cosmology experiments have direct sensitivity to DM particle physics \cite{Valluri:2022nrh,Mao:2022fyx,Dvorkin:2022bsc}. {Cosmological studies of DM should be supported as a key component of the HEP Cosmic Frontier program due to their unique ability to probe DM microphysics and link the results of terrestrial DM experiments to cosmological measurements.}

    \item {{The construction of future cosmology experiments is critical for expanding our understanding of DM physics.} Proposed facilities across the electromagnetic spectrum, as well as gravitational waves, can provide sensitivity to DM physics \cite{Chakrabarti:2022cbu}. The HEP community should make strategic investments in the design, construction, and operation of these facilities in order to maximize their sensitivity to DM physics.}

    \item Cosmic probes provide robust sensitivity to the microphysical properties of DM due to enormous progress in theoretical modeling, numerical simulations, and astrophysical observations. {Theory, simulation, observation, and experiment must be supported together to maximize the efficacy of cosmic probes of DM physics.}
\end{itemize}

\vspace{1em}
\noindent {\bf Particular Strengths and Capabilities}

\noindent Cosmic probes study DM on extreme large scales and in extreme environments that are inaccessible to terrestrial experiments.
As such, cosmic probes have sensitivity to DM properties over an extremely wide range of parameter space.

\begin{itemize}[nosep]

\item The SM of particle physics and cosmology can be tested at unprecedented levels of precision by measuring the cosmic distribution of DM. These measurements span an enormous range of scales from the observable universe to sub-stellar-mass systems
(e.g., the matter power spectrum, the mass spectrum of DM halos, DM halo density profiles, and abundances of compact objects)
\cite{Bechtol:2022koa,Bird:2022wvk,Brito:2022lmd}. Novel particle properties of DM (e.g., self-interactions, quantum wave features, tight couplings with radiation) can lead to observable signatures in the distribution of DM beyond the CDM prediction~\cite{Bechtol:2022koa,Drlica-Wagner:2022lbd}. The HEP community should support measurements of the DM distribution as a key element of its Cosmic Frontier program.

\item CDM makes the strong, testable prediction that the mass spectrum of DM halos extends below the threshold at which galaxies form $\mathcal{O}(10^7)~M_{\odot}$~\cite{Bechtol:2022koa}. Sub-galactic DM halos are less influenced by baryonic processes making them especially clean probes of fundamental physics of DM. 
We are on the cusp of detecting DM halos that are devoid of visible stars through several cosmic probes (e.g., strong lensing, the dynamics of stars around the Milky Way). The HEP community should pursue the detection of DM halos below the threshold of galaxy formation as an exceptional test of fundamental DM properties.

\item Extreme astrophysical environments provide unique opportunities to explore DM couplings to the SM that are inaccessible with conventional DM search experiments and span an $50$ orders of magnitude in DM particle mass~\cite{Berti:2022rwn}. Instruments, observations, and theorists that study extreme astrophysical environments should be supported as an essential means to constrain the expanding landscape of DM models and search strategies.

\item Numerical simulations of structure formation and baryonic physics are critical to robustly interpret observational studies and disentangle new DM physics from astrophysical backgrounds~\cite{Banerjee:2022qcb}, the necessary condition for addressing particle physics questions about the nature of DM. HEP computational expertise and resources can be combined with astrophysical simulation expertise to rapidly advance numerical simulations of DM physics.

\item The interdisciplinary nature of DM research calls for innovative ways of supporting a comprehensive pursuit of scientific opportunities cutting across traditional disciplinary boundaries. Sustained collaboration between particle theorists, gravitational dynamicists, numerical simulators, observers, and experimentalists are required to fully realize the power of cosmic probes of DM. Such large collaborations are naturally matched to the HEP community, but new mechanisms are also needed to support these emerging, interdisciplinary efforts.

\end{itemize}


\vspace{1em}
\noindent {\bf Complementarity}

\noindent Cosmic probes of DM lay the foundation for other direct experimental tests of DM particle properties, both within the Cosmic Frontier and across other experimental Frontiers. 
Cosmic probes expand the sensitivity to particle DM (CF01) and wave-like DM (CF02) to otherwise inaccessible ranges of mass and coupling strength.
Measurements of the distribution of DM are essential to interpret DM searches in terrestrial experiments (CF01 \& CF02) and indirect searches studying astrophysical systems (CF01).
Furthermore, many of the models studied by cosmic probes of DM have relevance in other frontiers.
For example, keV-mass sterile neutrinos  can be studied through their impact on cosmic structure formation (NF3).
As another example, self-interacting DM models are necessarily an indication of a non-minimal dark sector, which is complementary to dark mediator searches in accelerators (RF6).

Many of the experimental techniques used to make cosmic measurements of DM overlap heavily with experimental techniques used to measure dark energy, inflation, and cosmological neutrino properties (CF04, CF05, \& CF06), as well as indirect searches for DM interactions (CF01 \& CF07).
As such, the overlap with the Instrumentation Frontier in the development of low-noise photon detectors (IF02) and Computational Frontier in the development of numerical simulations (CompF2) and the implementation of machine learning for data analysis (CompF3), and the long-term interpretation and preservation of data (CompF7).


\subsection{Rare \& Precision Frontier 6 - Dark Sectors at High Intensities}
Intensity-frontier experiments offer unique and unprecedented access to the physics of low-mass DM.  Production of light DM is an especially powerful probe of thermal DM models, where the kinematics responsible for freeze-out is similar to the kinematics of accelerator-based particle production.  In addition, these experiments can produce other light, SM-neutral dark-sector particles that decay into SM particles, frequently with observably long lifetimes.  Discovery of such particles can shed light on the interactions, nature, and origin of DM.  Through these two complementary thrusts---also called out in the 2018 DM New Initiatives Basic Research Needs report (BRN) ~\cite{BRNreport}---accelerator-based experiments offer both excellent discovery prospects for light particle DM and the opportunity to characterize its particle properties. \vspace{1em}

\noindent \textbf{Scientific Opportunities \& Roadmap} 

The next decade offers opportunities for dramatic advances in dark sector searches, including orders of magnitude improvement in sensitivity to both DM production and long-lived particle decays, and thorough exploration of the milestone coupling ranges motivated by thermal DM across the MeV-to-GeV mass range. 
To realize these goals, the US high-intensity dark-sector community must do the following:
\begin{itemize}[leftmargin=1.0em]
\item \textbf{Exploit the capabilities of existing large multi-purpose detectors}, especially Belle II and LHCb.  These experiments can explore large regions of compelling parameter space for both GeV-scale DM and mediators to a dark sector that decay to ordinary matter.  
\item \textbf{Invest in the completion of the DMNI Program:} 
Dedicated small-scale experiments are required to explore scenarios that are not accessible to large multi-purpose detectors, in particular providing access to the feeble couplings motivated by sub-GeV thermal DM and to weakly coupled dark-sector particles with meter-scale lifetimes. 
Through the competitive DMNI program, the community and DOE selected two intensity-frontier projects, CCM200 and LDMX, to explore low-mass thermal DM. While CCM200 was completed in 2021 and is now operating, the LDMX missing-momentum experiment, like other DMNI-selected projects, still awaits project funding.  Completing the DMNI accelerator-based program offers a powerful opportunity at low cost to explore low-mass DM, improving on existing sensitivity by orders of magnitude and fully exploring the low-mass thermal DM region.
\item \textbf{Broaden the DMNI experimental portfolio to achieve the goals laid out in the DMNI Report:} The program selected for DMNI funds efficiently addresses the first thrust highlighted in the BRN report (DM production), but not the second (searches for visibly decaying dark-sector particles). Achieving the goals laid out in the DMNI report, and reaching the full potential of the accelerator-based dark-sector program, requires investment in experiments with complementary sensitivity.  A number of experiments have been proposed to explore the (semi)visibly decaying long-lived regime, and the US dark-sectors community should select which of these exciting ideas to fund in a second DMNI round. Regular DMNI-like funding opportunities will enable further expansion of the experimental program to provide complementary sensitivity, incorporate new ideas, and maximize discovery potential.
\item \textbf{Support dark sector theory} efforts to: better understand which dark-sector scenarios can address (current and future) open problems in particle physics; develop new ideas for exploring the dark sector;
and collaborate at every stage of new dark-sector experiments, from design through data interpretation. This type of theory work has been at the foundation of essentially all ongoing and planned 
experimental activities in this growing field. Support for theory-experiment collaborations and workshops will be important.
\end{itemize}

\vspace{1em}
\noindent \textbf{Particular Strengths and Capabilities} 

\begin{itemize}[leftmargin=1.0em]
    \item Intensity-frontier searches can systematically probe a broad range of simple, well-motivated dark sectors that are neutral under SM forces, including near-term coverage of thermal relic parameter space.  They can also explore generalized freeze-out scenarios such as strongly interacting massive particles and forbidden DM~\cite{Asadi:2022njl}. 
    \item In DM models where interactions are suppressed at low velocities, accelerator-based production is the favored path to detection because the non-relativistic interactions of cosmological and Galactic DM can suppress signals of its interactions by orders of magnitude.  Examples include inelastic (pseudo-Dirac or scalar) and Majorana models for light thermal DM.
   \item Intensity-frontier searches can probe DM couplings to a range of SM particles ({\em e.g.}\ hadrons, electrons, muons, photons) and detailed physics of the dark sector by discovering unstable dark-sector particles.
   \end{itemize}

\vspace{1em}
\noindent \textbf{Complementarity} 

The window on DM from Rare \& Precision Frontier experiments is both highly complementary to that of both Cosmic, Neutrino, and Energy Frontier probes, and multifaceted. 

Cosmic Frontier searches for light particle DM (CF01) include efforts targeting similar models of light DM. These experiments have complementary signal scaling, and hence carrying out both programs significantly increases discovery potential relative to either one on its own. In particular, direct and indirect detection are sensitive to DM interactions at non-relativistic velocities, while RF searches probe DM interactions at (semi)relativistic kinematics. The non-relativistic nature of direct detection can either enhance or suppress signals at a given interaction strength. Because the energies probed in RF experiments are comparable to the energies relevant for light DM thermal freeze-out, the range of production cross-sections expected for low-mass thermal relics is relatively compact---and accessible---regardless of the DM spin.  
Cosmic Probes of DM (CF03) also shed light on low-mass DM, and in particular on the DM self-interactions that are present in some dark-sector models and parameter regions.  

Neutrino Frontier searches (NF03) utilize the high intensity and high power proton beams together with the large scale, precision detectors to explore the intermediate mass range of DM (from a few 100s of keV to $<10$GeV) and pushing the lower bounds of the coupling strength both directly produced in the neutrino target and from the cosmic sources.  

Energy Frontier searches (EF10) employ similar techniques, but with a different model focus. In particular, EF experiments are needed to search for heavier DM ({\em e.g.}\ WIMPs), while generally having sensitivity to larger couplings, while RF experiments are optimized to search for the much weaker couplings typical of light DM models.   
In addition, a number of proposals have been made to co-locate experiments that will search for low-mass and long-lived dark-sector particles at EF facilities. 

%
%

\subsection{Energy Frontier 10 - Dark Matter at Colliders}

If particle DM signals are discovered via other frontiers, it will be imperative to produce and study these particles and their associated DM-SM interactions in the controlled conditions of high-energy accelerator-based experiments.
Conversely, high-energy experiments may discover evidence that new invisible particles are produced in collisions and may begin to study the nature of the interactions that govern their production, but evidence from other frontiers would be required to establish that the invisible particles are DM.
The existing and future program of DM searches at collider experiments is expansive, ranging from numerous signals of the canonical WIMP (featuring in supersymmetric theories), to lighter thermally-produced DM particles and the higher-mass mediators of their interaction, to extended dark sectors with complex, experimentally challenging phenomenology. 
More information can be found in the Energy Frontier report~\cite{Energy-Frontier-Report} and more-detailed topical group report~\cite{Bose:2022obr}.

\vspace{1em}
\noindent {\bf Scientific Opportunities \& Roadmap}

A chapter on the vision of the Energy Frontier can be found in Ref.~\cite{Energy-Frontier-Report} and can be summarized as follows: 

\begin{itemize}
\item In the immediate future, the Energy Frontier is delivering the High Luminosity LHC (HL-LHC), a leading P5 priority from the last Snowmass.
  In the next decade, upgraded experiments at the HL-LHC can discover or exclude thermally produced, kinetically-mixed DM above the electroweak scale (about 100 GeV). These experiments will also significantly tighten constraints on DM particles coupling to the Higgs boson and test supersymmetric and other particle DM candidates that have been long-term targets of the field. 
\item In the intermediate future, an electron-positron collider can further explore the hypothesis that DM is created via a Higgs boson or other beyond the Standard Model (BSM) portal particles, especially DM models that favor couplings to leptons, and observe early signals of new particles that could be identified as DM or DM mediators directly or in precision measurements. 
\item In the longer term, a discovery machine (hadron or muon collider) will allow direct exploration at far-higher energy scales for dark-sector physics than will be possible at the HL-LHC. Such a machine has the greatest prospects to discover new fundamental particles as part of WIMP multiplets, and it will reach the thermal milestone for kinetically-mixed thermal DM from 3 GeV of mass, complementing the lower-mass Frontier experiments.  
\end{itemize}



\vspace{1em}
\noindent {\bf Particular Strengths and Capabilities}
\begin{itemize}
\item Collider experiments probe DM in a controlled environment over a range of energies. The high-energy of the collision allows access to lighter, relativistic DM and other invisible particles (e.g., neutrinos), as well as to heavy (TeV-scale) DM; 
\item Collider searches are not restricted by the characteristics of DM that already exists in the halo, and therefore they aren't limited by low cosmic event rates or subject to astrophysical uncertainties; 
\item Collider experiments may be able to probe effective DM couplings to a range of SM particles, as well as the DM flavor structure, and other particles associated with the dark-sector physics, such as interaction mediators or unstable dark sector particles. 
\end{itemize}

\vspace{1em}
\noindent {\bf Complementarity}
\begin{itemize}

\item A simultaneous discovery of DM microphysics at a Cosmic Frontier experiment and at a high energy collider would combine the capabilities of both types of experiments to constrain the type(s) of DM particles present and their interactions. Similarly, hints from one of the two experimental approaches would focus and refine the efforts of the other. This is exemplified in \textit{Case Study A - Minimal WIMP} (Section \ref{sub:minimalWIMP}). 

\item For types of DM for which this sort of simultaneous discovery would be difficult, collider experiments provide complementary sensitivity to direct detection experiments, such as when direct-detection signals are suppressed at low velocities. 

\item Future colliders would produce thermal DM in a complementary mass range with respect to other accelerator experiments; both would be necessary to span the entire mass reach for thermally-produced, kinetically-mixed DM. 

\item In the context of certain low-mass portal DM models targeted by the Cosmic and Rare Processes and Precision Frontiers, colliders can search for the high-mass particles predicted in order to establish theoretical consistency of the models, as mentioned in the EF10 report \cite{Bose:2022obr}. 

\item High-energy collider beams can be used for co-located DM experiments searching for long-lived, dark sector particles. An example is the case of the proposed Forward Physics Facility \cite{Feng:2022inv,Anchordoqui:2021ghd,Batell:2021blf} for the HL-LHC, but similar facilities would provide similar capabilities at other future colliders. 

\item To help each experimental frontier understand where to look for particle DM, the theory frontier is vital. One must assume one or more theoretical models of DM to relate information from multiple experiments to each other, but there is a vast set of possible models and, until signals are observed, few clues as to which models are nearer to reality. The theory is especially important for collider searches, as the model details can greatly affect the kind and size of signals sought. 
Efforts are needed to improve sharing of results across frontiers and to allow more easily testing models against the results (e.g., with global fits, with a recent summary of results from one of these tools in \cite{White:2022dsq}).
\end{itemize}

\subsection{Neutrino Frontier 3 - Dark Matter in Neutrino Experiments}



Neutrino experiments, while primarily aimed at understanding the properties of neutrinos, can also provide sensitivity to dark matter. DM is intimately tied to the neutrino sector in many well-motivated frameworks: it can, for instance, be a right-handed (sterile) neutrino \cite{Drewes:2016upu,Shakya:2015xnx,Asadi:2022njl} or be part of a broader sector that couples to the SM through a neutrino portal. Sterile neutrinos can be realized across many mass scales; the vacuum mixing of sterile neutrinos with active states is constrained by experiment, astrophysical, and cosmological considerations \cite{Kusenko:2004qc}.
Short baseline oscillation experiment (e.g., MiniBooNE \cite{MiniBooNE:2018esg}) and reactor neutrino \cite{DayaBay:2017jkb,Hayes:2017res} anomalies have been interpreted as active-sterile oscillation with large mixing at the $\sim$1 eV mass scale \cite{Acero:2022wqg}.
Likewise, discordance in measurements of the Hubble parameter in the early- and late-time Universe \cite{Planck:2018vyg,Riess:2019cxk,Freedman:2019jwv,Wong:2019kwg} has led to many ideas invoking neutrino-sector BSM considerations.
A more comprehensive discussion of the synergies between cosmological and laboratory searches for neutrino physics can be found in \cite{Abazajian:2022ofy}.

More generally, since most SM particle decays produce neutrinos, any DM interaction with SM particles will inevitably give rise to a signal that can be probed with neutrinos. Furthermore, an improved understanding of neutrino fluxes, properties and cross sections can also assist in understanding the backgrounds in DM searches with neutrino experiments and in other DM experiments. An example of the latter is the need to understand the neutrino fog in direct detection searches.     

Typically, DM searches in neutrino experiments manifest in the direct and indirect search categories.  
In direct searches for DM interactions or decays, DM can be produced in both natural or artificial sources.
Generally, DM candidates are nonrelativistic, and the particles produced via DM interactions are below the detection threshold of neutrino experiments.
However, in ``boosted'' scenarios, annihilation or decay of the non-relativistic DM in DM concentrated regions, such as the Sun or the Galactic Center, can produce lighter, relativistic DM with small flux and (semi)relativistic kinematics that can be detected by massive, underground neutrino detectors \cite{Berger:2022cab, Batell:2022xau}.

In parallel, the accelerator complex used to produce neutrino beams is essentially a high-intensity proton beam with a fixed target, which allows us to probe dark sector particles weakly coupled to the SM.
Together with advanced reconstruction algorithms, sophisticated analyses can be conducted to look for DM or dark sector candidates.
Similarly, nuclear reactors produce intense neutrinos and can potentially be sources of such particles.

On the other hand, DM can be indirectly probed or constrained by anomalous spectra of neutrino fluxes,
such as supernova neutrinos, atmospheric neutrinos, and neutrinos produced by artificial sources.
The anomalies include DM-induced neutrinos, and other sophisticated models.
\vspace{1em}
 


\noindent {\bf Scientific Opportunities \& Roadmap}

\begin{itemize}
\item {\bf Realization of DUNE in its full scope within the next decade} will expand the opportunities in both direct and indirect searches of DM and dark sector particles through the high intensity, high power proton beams as well as the powerful near detector and the large scale, high precision, low threshold far detectors.
\item {\bf Advancing the understanding of neutrino backgrounds} from natural or artificial neutrino sources.  As neutrino background often exists in DM searches and is challenging to separate from the DM signal, it is important to precisely characterize its flux and the modeling of its interactions with the nuclei.  Small-scale supporting experiments, such as those in collaboration with the nuclear physics community, may be required to conduct precise measurements.  The community also benefits from such experiments on the related training and R\&D opportunities, as well as promoting and strengthening much needed close, synergistic collaboration with the nuclear physics community.
\end{itemize}

\vspace{1em}
\noindent {\bf Particular Strengths and Capabilities}
\begin{itemize}
\item Neutrino experiments are the natural places to search for DM by looking for DM-induced neutrinos, such as neutrinos from DM decays or those produced via DM annihilation. The broad energy range of neutrinos produced by natural sources provide probes of DM-induced neutrinos.
\item Neutrino detectors based on different technologies are sensitive to complementary neutrino flavors and energy ranges, which cover a large parameter space of DM-induced neutrinos and (boosted) DM.
\item Massive, underground neutrino detectors offer a large number of target elements for detecting (boosted) DM with relatively small fluxes. The underground locations also significantly reduce the cosmic ray background.
\item Accelerator-based neutrino experiments have intense proton beams impinging a fixed target. The modern near detectors (or detectors at short baseline) typically have high resolution in energy and time, as well as the capabilities of calorimetry and superb particle identification. This allows not only precision measurements of neutrinos, but also searches for DM, in particular in the intermediate mass range bridging between EF and CF.
\item New technology developed for neutrino experiments, such as LArTPCs used in DUNE, provides unprecedented capabilities, such as millimeter spatial resolution and better than $1^{o}$ angular resolution for tracking, and a few tens MeV of proton detection threshold. This significantly improves sensitivity to, for example, boosted DM in its elastic and in-elastic scattering channels. 

\end{itemize}

\vspace{1em}
\noindent
{\bf Complementarity}
\begin{itemize}
\item By precisely measuring neutrinos fluxes from natural sources, neutrino experiments can uniquely contribute to the landscape of DM indirect detection of DM-induced neutrinos.
\item Underground, massive neutrino detectors are uniquely sensitive to direct detection of boosted DM through their elastic and inelastic scattering channels, which is complementary to searches in direct DM detection experiments. In addition, neutrino detectors explore kinematic phase space of DM that is complementary to those probe by the Energy and Cosmic Frontiers.
\item It is important to note the complementarity within the Neutrino Frontier -- complementary information is offered by neutrino experiments based on different technologies, highlighting the importance of a program with a healthy breadth.
\item Accelerator-based neutrino experiments can conduct complementary direct searches for sub-GeV DM from the intense proton beam. This is in synergy with the Rare \& Precision Frontier, and the sub-GeV DM searches are typically complementary to the sensitive ranges of the searches in the Energy and Cosmic Frontiers.

\end{itemize}

\subsection{Theory Frontier 9}


Theory is the language by which the results of different experiments can be compared. 
A theory of dark matter is a rigorous mathematical framework to quantify the space of experimental possibilities, see~\cite{Green:2022hhj}. 
%
%
Broadly, theory contributions play the following complementary roles in the quest for dark matter:

\begin{itemize}

\item \textbf{Define connections between different experimental programs and Frontiers:} 
A theoretical model of DM defines the ways in which experimental probes across different frontiers are related. In this way, theory is the \emph{glue} of dark matter complementarity: it allows us to connect data from different experiments to exclude a particular scenario or plan specific searches to verify a potential discovery.

\item \textbf{Motivate specific experimental directions:} Theory can motivate experimental programs for DM searches. Of the vast number of allowed dark matter models, the most attractive ones tie DM to some deeper aspect of particle physics that is known to us, connecting the DM question to other questions that might be the primary focus of various Frontiers. For instance, the hierarchy problem inspired the WIMP DM paradigm, connecting DM to the central goals of the Energy Frontier. Likewise, models of neutrino mass generation have inspired sterile neutrino DM, and more broadly, neutrino portal DM, establishing connections with the Neutrino Frontier. Similarly, the strong CP problem gave birth to axion DM. 

\item \textbf{Identify DM detection capabilities of experimental programs:} As technological progress and new experimental insights broaden the ways we can search for DM, it becomes important to identify plausible DM scenarios and properties that can be probed with such approaches, which can further sharpen the goals of such experimental programs. A recent example of such interplay is the development of the dark sector program in the rare processes Frontier. 


\end{itemize}

\subsection{Instrumentation Frontier}
Advances in instrumentation support every aspect of the hunt for dark matter, with new technologies often opening up new regions of parameter space for exploration. Two examples coming from direct detection experiments are in quantum sensors and noble element detectors. The rapid progress in quantum sensors over the past decade have been key in the success of the wave-like dark matter program. Such sensors come in a wide range of technologies: atom interferometry, magnetometers, calorimeters, and superconducting sensors to name a few. In searches for WIMP-like dark matter, the development of liquid noble detectors has paved the way for the huge strides in sensitivity seen since the last Snowmass, with Liquid Xenon detectors leading the way with Liquid Argon close behind. Of course, there are many more examples that enable dark matter searches in all the frontiers, including development in photon sensing, timing, calorimetry, etc. Continued R\&D into instrumentation, development of the technical workforce, and tools to share common knowledge will be an important component of the future dark matter program. More information can be found in the Instrumentation Frontier report~\cite{IFReport}. 

\subsection{Computing Frontier}

Computing is a critical component of any search for dark matter, from the collection and storage of raw data, through various stages of data movement, data processing and data analysis, all the way to the interpretation of results. 
Interestingly enough, the majority of dark matter searches are converging towards comparable scales in data volume~\cite{SmallExperiments,CF1-WG4-Report}. For example, direct detection experiments in the CF are approaching a raw data throughput of order 1 PB/year. 
While not an unusual scale for large current EF or NF experiments, or even CF surveys, 
these volumes can present a significant challenge in the direct detection community which has not historically prioritized the development of a scalable computing infrastructure in support of its scientific ambition. 
Likewise, future EF experiments will see a rapid increase in needs for both storage and computing power towards exascale datasets that is not matched by technological developments and budgets~\cite{HEPSoftwareFoundation:2017ggl}. 


Given this convergence in data size increase and complexity, some common computing needs and themes can be identified to enable successful dark matter searches across all frontiers, including: 

\begin{itemize}
\item build stronger partnerships with the national supercomputing facilities, by lowering the barrier of entry and providing common tools and shared engineering. Solicit community input on architecture evolution, and facilitate the use of heterogeneous computing resources;
\item support scalable software infrastructure tools across communities, avoiding duplication of effort. These tools run the gamut of data management and archiving, event processing, reconstruction and analysis, software management, validation and distribution;
\item enhance industry collaborations on machine learning techniques and provide access to external experts. Foster efforts to understand uncertainties and physical interpretation of machine learning results.
\end{itemize}

Simulations and modeling are essential to all dark matter searches, and vital at every stage of an experiment's life-cycle, from instrument conception to interpretation of the results. It is essential to have an efficient, well-maintained, well-understood and thoroughly validated simulation infrastructure, encompassing a variety of components from cosmological modeling, to event generators and detector simulation frameworks. The simulation needs for much of the dark matter community reflect the CompF consensus~\cite{CompF-Report}, namely: continuation of Geant4 support and training; continued support for event generators, including those developed as part of a national security program; support for detector-specific simulation packages; and enhanced opportunities for cross-collaboration communication. While cosmic surveys require a different type of simulation (large cosmological volumes), many of the infrastructure needs are complementary \cite{Banerjee:2022qcb,Alvarez:2022rbk}. 

Finally, an additional challenge specific to complementarity is the need to exchange data between different experiments and even frontiers, which implies the necessity to converge on data formats and analysis tools. The dark matter community should take the lead in advocating for widely-adopted data and software standards, in support of global analyses of experimental results. Complementarity studies would be drastically simplified by an Open Data paradigm, with the added benefit of enhancing the credibility of our potential discoveries, since progress in the field of dark matter will require thorough scrutiny.

\subsection{Underground Facilities}
Direct dark matter experiments must be sited in underground facilities to evade cosmic ray backgrounds, and multiple new underground dark matter experiments are expected and being planned (at both large and small scales). Large neutrino experiments also require space underground. Currently, underground facilities are largely subscribed by existing projects, with only limited space available in the coming years. There is, then, a clear need for additional underground space, tailored to the needs of neutrino and dark matter experiments.
This underground space must accommodate experiments across scales, including large liquid noble or freon experiments and smaller installations, for example mK facilities. Assembly of future experiments will occur largely in the underground environment, requiring underground radon-free clean rooms. Given the volume of gas/cryogen, future liquid noble experiments also require underground areas for staging (e.g., gas storage) and experiment utilities (e.g. pumps, distillation). These new suitable spaces must be available by the late 2020s to meet the demand, which may be met in North America by proposed new excavations at SURF or SNOLAB. More information can be found in the UF report.

\subsection{Accelerator Frontier}
Current and future accelerator facilities are the underpinning of both intensity-frontier and energy-frontier searches for dark matter and dark sectors. Several beam facilities for axion and Dark Matter (DM) searches are shown to have great potential for construction in the 2030s in terms of scientific output, cost and timeline, including PAR (a 1 GeV, 100 kW PIP-II Accumulator Ring); in general, we should efficiently utilize existing and upcoming facilities to explore dedicated or parasitic opportunities for rare process measurements - examples are the SLAC SRF electron linac, MWs of proton beam power potentially available after construction of the PIP-II SRF linac, spigots of the future multi-MW FNAL complex upgrade, and at CERN, a Forward Physics Facility at the LHC, etc.  At the energy frontier, an e+/e- Higgs Factory (e.g. FCC-ee, C3, etc) will likely be the next major accelerator facility while interest in discovery machines such as O(10 TeV c.m.e.) muon colliders has also gained significant momentum. In all cases, these machines’ discovery capabilities are open-ended, and certainly include significant territory for dark matter discovery. To be in a position for making decisions on energy frontier collider projects viable for operation in the 2040s and beyond at the time of the next Snowmass/P5, these concepts could be explored technically and documented in reports by the end of this decade.  More information can be found in the AF report~\cite{AF-Report}.

\subsection{Community Engagement and Workforce Development}

The last two decades have seen an explosion in the number of physicists engaged in dark matter detection, and the enthusiasm for the topic has been palpable throughout Snowmass 2021. In order to support this enthusiasm, the community needs to enhance its engagement efforts at all levels of society, most notably with the education system, industry partners, and policy makers. Outreach to stakeholders includes not only those we directly partner with in industry, education, and government, but also broadly communicating with the general public about our work. Therefore, our community -- and broader society -- will benefit from material support for these kinds of outreach activities, which are core, not peripheral to our work. 

Outreach is both about widening access to information and also about recruitment. Helpfully, the exciting mystery of dark matter and its connections to the most basic questions about our universe not only motivate its study but also act as a powerful recruiting tool for HEP. As is evident by this review, the search for dark matter is highly interdisciplinary and is therefore an excellent training ground for our next generation of scientists. In dark matter searches, quantum sensing and AI/ML are critical for extracting these elusive signals, so this training is not only aligned with the needs of HEP but also well-aligned with US priorities as a whole. 

Career pipeline and development are crucial to sustaining such an expansion, which in turn requires a renewed focus on the diversity, equity, inclusion, and accessibility of the field. The historical exclusion of marginalized people from high energy physics is fundamentally harmful to the humanity of individuals who are excited and curious about science. Moreover, creating equal opportunity and equality in particle physics is essential to professional success in our field, ensuring breadth of perspectives and a deep talent pool. Developing a broad talent pool sustains both scientific advancements in high energy physics and the democratic principle of publicly supported activities that are by and for the people.  More information can be found in the Community Engagement Frontier report~\cite{CEF-Report}.

Ensuring democratic access to opportunities in HEP goes hand in hand with providing regular training opportunities. For the HEP workforce specifically, the move to larger facilities and projects has led to larger gaps between the design and commissioning of new efforts, which can lead to leaks in the pipeline of key knowledge holders. The search for dark matter provides many small and medium-scale projects that allow scientists to take part in all phases of the experiment from design to commissioning and analysis. For this reason, it is critical that HEP supports a portfolio of small and medium-scale projects for both their discovery potential and for training of its workforce. For large-scale projects such as future colliders at the Energy Frontier, involvement in the long-term design and R\&D toward these projects uniquely provides training in hardware development and construction that supplements that obtained in the operation and data analysis of experiments already underway. This broad experience is vital to the development of detector capabilities and analysis methods that can capture the atypical, difficult-to-observe signals of dark matter in these large, general-purpose experiments.


\section{Case Studies}
\label{sec:CaseStudies}

\begin{figure}[htp]
\begin{center}
\includegraphics[width=\textwidth]{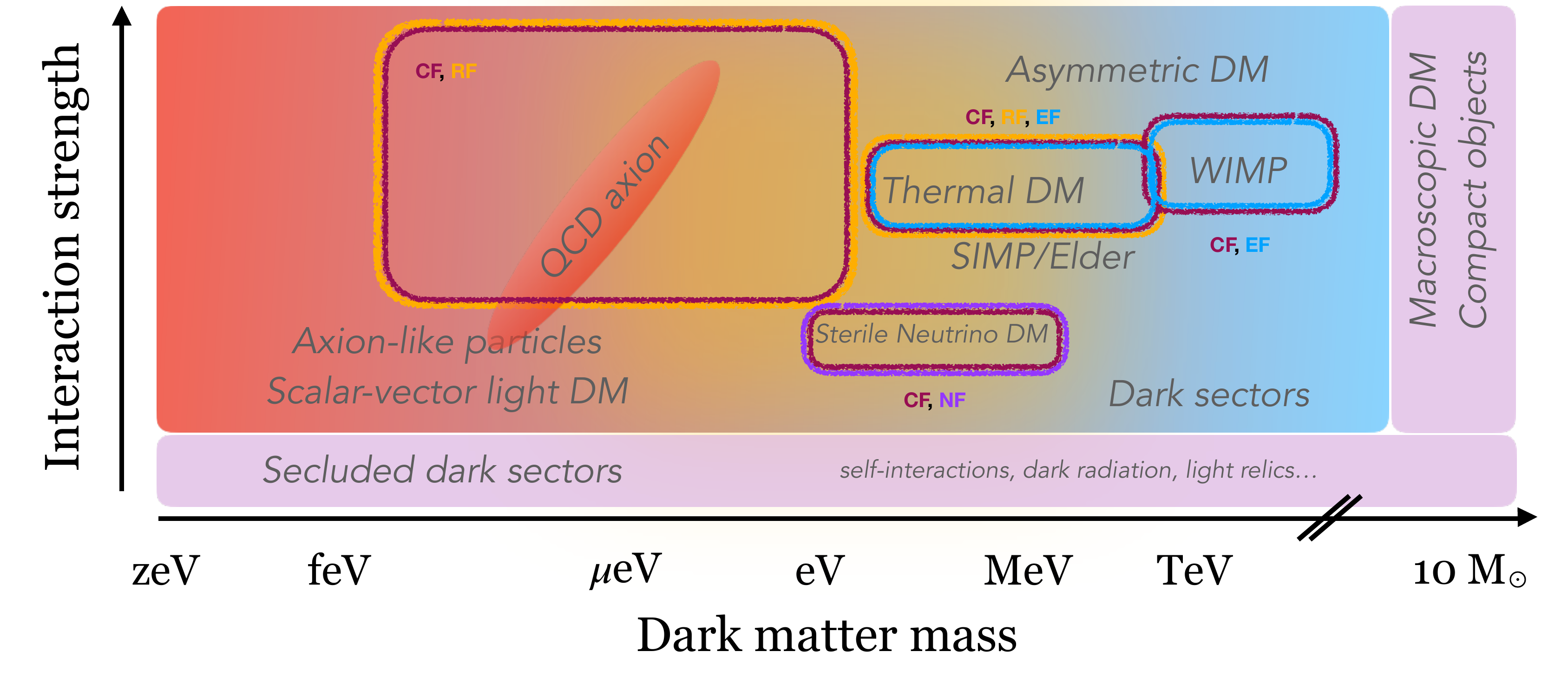}
\end{center}
    \caption{Summary of case studies presented in this document, shown in the context of a sketch of the coupling-mass plane including the parameter space typical of some of the rich variety of DM theories possible. The rounded rectangles highlight the classic minimal WIMP paradigm, vector-portal dark matter (e.g. DM-SM interactions mediated by a $Z^\prime$ or dark photon), sterile neutrino dark matter, and wave-like (axion) dark matter examples discussed in subsequent sections. The shaded colors in this sketch are suggestive of the Frontiers with experiments represented in the case studies in a given region, with color coding specified near the rounded rectangles.}
    \label{fig:allcasestudies}
\end{figure}

Fig.~\ref{fig:allcasestudies} provides a graphical summary of the breadth of theoretical scenarios that can provide dark matter candidates. The  possibilities span enormous ranges in dark matter mass 
and interaction strength.

In the case studies below, we discuss several scenarios in depth to illustrate how complementarity between DM searches could enable discovery of the fundamental nature of DM and allow triangulation of its properties. We begin with a detailed case study of a realization of the classic WIMP paradigm, where the complementarity between the Energy Frontier and Cosmic Frontier has been relatively well-studied, and then summarize important aspects of complementarity in a number of examples spanning a broader range of DM scenarios. Finally, we provide a very brief outline of other complementarity case studies from recent reports.

\subsection{Minimal WIMP dark matter}
\label{sub:minimalWIMP}


The first case study used to depict DM complementarity across Frontiers is a minimal, concrete realization of the canonical WIMP paradigm.

The DM particle can be embedded within a new particle multiplet that interacts with SM particles via the weak interaction \cite{Cirelli:2005uq}. Accidental or imposed symmetries lead to the stability of the lightest particle of this multiplet, providing a suitable DM candidate that can be produced thermally and satisfy relic density constraints. 
Such multiplets are foreseen in electroweak supersymmetric theories, as superpartners of the SM gauge/Higgs bosons, and possess (predictive) small cross-sections for detection that are consequently not fully probed by current experimental efforts. 
We will focus on two widely studied scenarios where the DM particle is part of a Dirac fermion doublet (called the Higgsino) or of a Majorana fermion triplet (Wino)\footnote{Many of the same general principles also apply to larger electroweak multiplets; for a discussion of other options see e.g. \cite{Bottaro:2021snn, Bottaro:2022one}.} 
The advantage of such scenarios is their predictivity: the only free parameter is the mass of the Wino and Higgsino, and this can be fixed by requiring that they constitute all the measured DM relic density. 
This leads to TeV-range mass predictions for Wino and Higgsino as thermal DM candidates, 1.1 TeV and 2.8 TeV respectively. 
If non-standard or non-thermal production mechanisms are at play, then the DM particles can also be lighter \cite{Cirelli:2007xd}. 

Given the low cross-sections and heavy masses of these DM candidates, indirect detection experiments (as well as large future direct detection experiments) are well suited for their identification. The Wino and Higgsino can also be produced and observed at future colliders. Predicting the observational signatures of these candidates, and their thermal history in the early universe, can require sophisticated theoretical techniques to capture the effects of long-range interactions due to SM gauge boson exchange \cite{Hisano:2004ds} and loop diagrams enhanced by Sudakov logarithms \cite{Baumgart:2014vma, Bauer:2014ula, Ovanesyan:2014fwa, , Baumgart:2018yed, Beneke:2019gtg}.

\paragraph{CF: Indirect detection} The annihilation of two WIMPs into photons and gauge bosons produces a characteristic signal in the energy spectra detected by gamma-ray telescopes and other indirect detection experiments. Exchange of SM gauge bosons provides a long-range force between DM particles that can enhance the annihilation cross section above the usual thermal value, and in particular enhances the branching ratio to produce a gamma-ray line at the DM mass \cite{Hisano:2004ds}.

Existing limits on gamma rays from the inner Milky Way using air Cherenkov telescopes already tightly constrain thermal Wino DM, although these bounds weaken when taking into account uncertainties on the DM density distribution toward the Galactic Center \cite{Rinchiuso:2018ajn, Hryczuk:2019nql}, motivating probing this scenario further in the near- and medium-term.\footnote{Observations of dwarf galaxies provide a lower-background target compared to the Galactic Center, with smaller uncertainties on the DM density profile, but also a lower signal. In an example of successful cross-community collaboration, Ref.~\cite{Hess:2021cdp} uses data from multiple gamma-ray telescopes and common statistical tools to set stringent bounds on DM annihilation from dwarf galaxies, in a broad mass range from 5 GeV to 100 TeV.} Antiproton limits on Wino DM are also competitive \cite{Cuoco:2017iax}, motivating studies to reduce the uncertainties on cosmic-ray production and propagation.



Future indirect detection experiments (e.g. CTA) will be able to reach the Higgsino thermal target \cite{Hryczuk:2019nql}. For larger multiplets, the formation of bound states becomes important (e.g.~\cite{Mitridate:2017izz, Mahbubani:2020knq}), and further theoretical study of predicted signatures may be required. An initial detection would likely involve observation of gamma rays or charged particles close in energy to the DM mass, but lower-energy observations with future detectors could reveal spectral line signatures from bound state formation and transitions, permitting DM spectroscopy \cite{Mahbubani:2020knq}.

\paragraph{CF: Direct detection} Direct detection signatures of the Wino and Higgsino in DM-nucleon interactions are expected to be very rare, and therefore require large exposure for detection. The nominally leading-order diagram is 1-loop and involves W boson exchange, suggesting a naturally sub-weak-scale cross section that combines with the large DM mass to render detection difficult; furthermore, there is a generic cancellation between contributing amplitudes that further reduces the expected cross section \cite{Chen:2019gtm}. 


Large future direct detection experiments such as DARWIN can probe the Wino scenario, with cross-sections still above the neutrino fog \cite{Chen:2018uqz}. 
For the Higgsino case, the direct detection cross-section for small mass splitting between the lightest and next-to-lightest SUSY particles is below the neutrino fog, likely requiring novel approaches. 

\paragraph{EF: Future colliders} Electroweak multiplet DM at colliders can be detected in three ways (1)  via missing transverse momentum searches, where the WIMP (invisible to detectors) recoil against one or more visible SM particles; (2) from the decay of long-lived intermediate states into a charged particle and the WIMP, leading to a “disappearing track” signature in the detector (3) from loop effects due to the presence of the new particles \cite{DiLuzio:2018jwd}. 
The first two signatures require TeV-scale particles (Wino and Higgsino, but also other particles in the cascade) to be produced in highly energetic collisions, while quantum loop effects can be detected in precision measurements at lower center-of-mass energy colliders.

In the near-term (High Luminosity LHC), non-thermal WIMPs with masses below the TeV can be detected at colliders using the disappearing track signature.
Precision measurements at future lepton colliders and Higgs factories, to be built in the next NN years, can reach masses of 0.3-0.5 GeV. 
In the longer term, a multi-TeV lepton or hadron collider can meet the thermal target for both Wino and Higgsino (and eventually even for larger multiplets). 

More information on the projections for future colliders can be found in Section XI of the Beyond-the-Standard-Model report of the Energy Frontier~\cite{Bose:2022obr}. 

\begin{figure}[htp]
\centering
\subfloat[]{\label{subfig:WIMP_ID_Colliders}\includegraphics[width=0.7\textwidth]{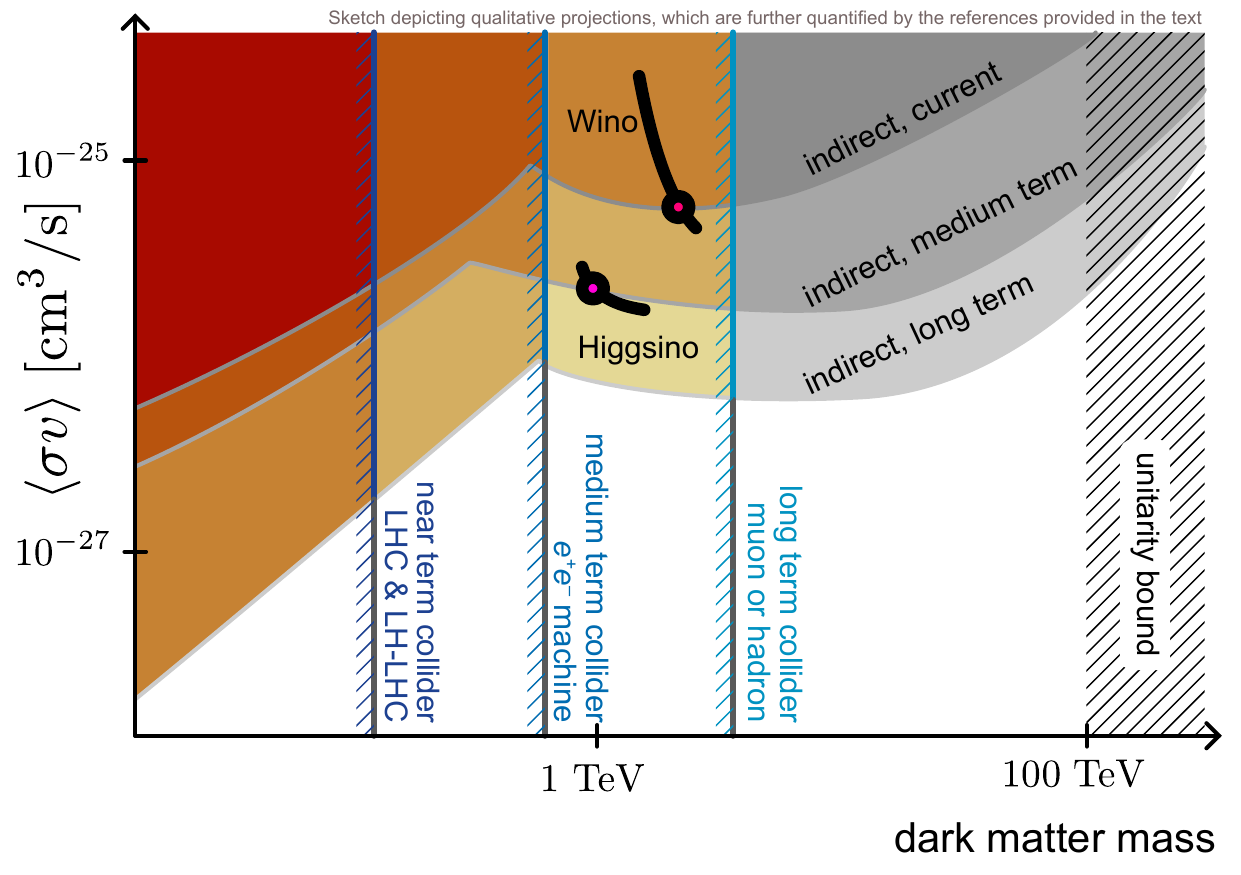}}\\
\subfloat[]{\label{subfig:WIMP_DD_Colliders}\includegraphics[width=0.7\textwidth]{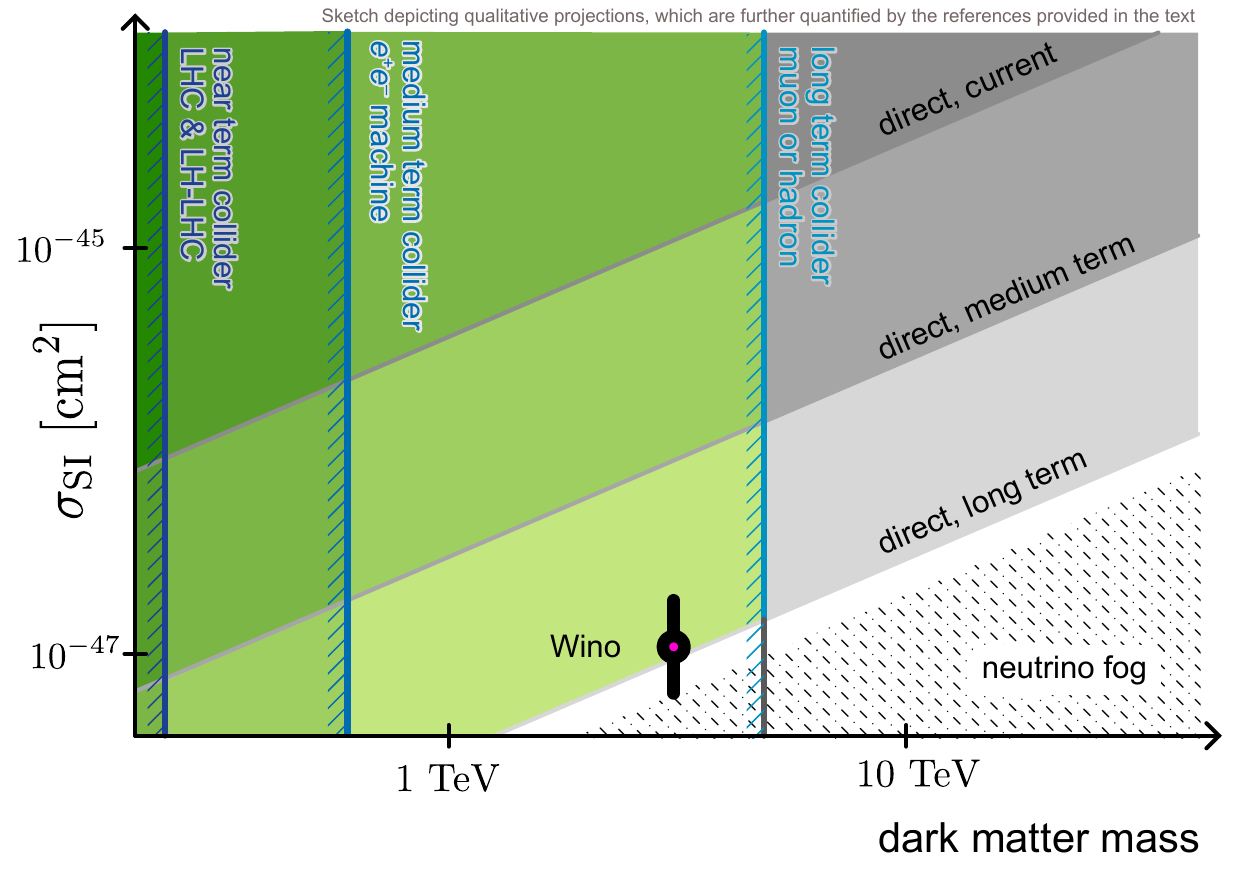}}

\caption{Indicative sketches depicting qualitatively how future collider and indirect detection (\protect\subref{subfig:WIMP_ID_Colliders}) or direct detection (\protect\subref{subfig:WIMP_DD_Colliders}) may complement each other during a discover of Wino or Higgsino DM. The qualitative coverage shown is further quantified by the references provided in the text. Each figure sketches two types of constraints representative of those in Refs.~\cite{Energy-Frontier-Report,Bose:2022obr,Boveia:2022jox,Strategy:2019vxc}: the present experimental coverage of the annihilation of spin-independent WIMP-nucleon cross-section as a function of WIMP mass, as well as the projections for medium- and long-term experimental proposals described in the references. Regions of overlapping coverage, where complementary observations in both types of experiments would be possible, are indicated by saturated colors. Regions accessible by one of the two types of experiments only are shown in muted colors or grayscale. Solid points indicate approximate targets for Wino and Higgsino DM, as discussed in the text. Also shown are regions at lower cross-section values where neutrino interactions with direct detection experiments contribute background to a search.}
    \label{fig:WIMP}
\end{figure}

\paragraph{Complementarity highlights} 
Future collider, direct and indirect detection experiments are probing the same parameter space for Wino and Higgsino models. 
This can be seen visually from the sketch in Fig. \ref{fig:WIMP} which reprises results from Refs.~\cite{Cooley:2022ufh,Bose:2022obr}.

A discovery of a signal with a cross-section compatible with a minimal WIMP (Higgsino or Wino, with masses around the TeV scale) in direct and indirect detection experiments would be complementary to evidence of the same particle at next-generation colliders, and if such a discovery comes during the planning phase of such colliders, provide crucial input to the collider design. 
Colliders will bring further information on the WIMP’s interactions with SM particles and its associated particle spectra. 
Even if no signal is observed in this mass range, this complementarity can be extended to the multi-TeV range in higher-dimension multiplets, with indirect and direct detection experiments contributing to the planning of future generations of colliders. 

Simultaneously, accelerator experiments can furnish improved measurements of the production rate for antiprotons and other antinuclei, relevant to cosmic-ray signals of DM annihilation (see \cite{CF1WP5} for a more in-depth discussion). These measurements will help reduce systematic uncertainties in both the signal and background for cosmic-ray antiprotons, rendering antiproton constraints on the Wino parameter space more robust.

An example timeline could involve a first hint of events in the full datasets (available late 2020s) of LZ and/or XENONnT, signaling high-mass DM with an interaction cross section comparable to that predicted by the Wino or DM inhabiting a larger electroweak multiplet. Such a direct detection signal would be of profound interest, and would probe the local velocity distribution of DM, but would not significantly constrain the DM mass or the details of how it interacts with the SM. A subsequent confirmation in the DarkSide-20k experiment, using argon as a target material rather than xenon, would constrain the DM couplings to the SM, while next-generation experiments such as XLZD/Argo/Darwin could provide higher significance for the measurement of DM properties.

On the same timescale, SWGO or CTA could identify a gamma-ray indirect signal from the Galactic center and/or dwarf galaxies, and measurements at NA61/SHINE, LHCb, and ALICE (see \cite{CF1WP5} and references therein) could enable more precise theoretical predictions for antiproton cosmic-ray signals, allowing detection of any discrepancy between theoretical predictions and AMS-02 observations of the cosmic-ray antiproton spectrum. Comparison of gamma-ray signals from various targets (e.g. dwarf galaxies and the Galactic Center) and any antiproton signal would constrain the distribution of DM throughout the Galaxy and the branching ratios for annihilation into different SM particles, and detection of a gamma-ray line signal would tightly constrain the DM mass. A search at a next-generation collider with sensitivity to multi-TeV electroweak DM (in the 2040s) would then be required to nail down the couplings of DM to SM particles, complementing the  data from the indirect and direct detection signals, and identify the other particles in the multiplet.

Alternatively, null results at collider experiments could significantly constrain the interpretation of a putative DM signal from direct or indirect detection. A recent example of this occurred in the context of the Galactic Center excess (GCE), which could potentially be explained by thermal relic DM with a tens-of-GeV mass. 
Null results at collider and direct detection experiments set stringent bounds on effective operators and simplified models that could generate the GCE (e.g.~\cite{Alves:2014yha,Berlin:2014tja,Dolan:2014ska,Abdullah:2014lla,Agrawal:2014una}).

\FloatBarrier

\subsection{Generic Beyond the Standard Model (BSM)-mediated and vector portal dark matter}
\label{sub:BSMMediation}

In case of a signal in a Cosmic Frontier experiment (e.g. direct or indirect detection), it will be necessary to pinpoint its interactions and possible related particles in order to characterise the DM sector. 
Different types of cosmic probes and target materials can already shed some light on the kinds of interactions undergone by the DM particle. 
However it is by producing the same kind of DM in the lab, where the initial state is known (in terms of colliding particles, or beam particles impinging on a target), that we will be able to pinpoint which interactions and processes the DM particle has with ordinary matter. 
Likewise, a signal indicating the presence of an invisible particle in a fixed target or collider experiment can  be interpreted in terms of DM benchmarks.
Nevertheless, it is through a simultaneous discovery in Cosmic Frontier experiments that we will be able to ascertain the cosmological nature of the DM candidate. 

This kind of complementarity can be illustrated using theoretical scenarios that extend the WIMP paradigm to include an additional particle beyond the SM.
This new particle mediates the interaction between DM and SM with a coupling strength that is comparable to or weaker than the coupling strength of the weak interaction.
These mediator particles can decay into both DM and SM particles, offering further insight into the DM-SM interaction through experiments that have been designed to discover new particles decaying visibly as well as into invisible particles, such as those at colliders and accelerators.  
A thermal history for the DM candidates in the early universe can be attained depending on the coupling types and strengths of the mediator (or \textit{portal}) particle, as well as on the mediator mass and on the mass of the DM particle. 
Specific realizations of these models are used as benchmarks by the collider and accelerator communities, see the generic BSM-mediated benchmark models in Refs. \cite{Abercrombie:2015wmb,Energy-Frontier-Report} and the vector portal DM models in Refs. \cite{Beacham:2019nyx,Gori:2022vri}. 

\begin{figure}[htp]
\begin{center}
\subfloat[]{\label{subfig:BSMMediationSI}\includegraphics[width=0.7\textwidth]{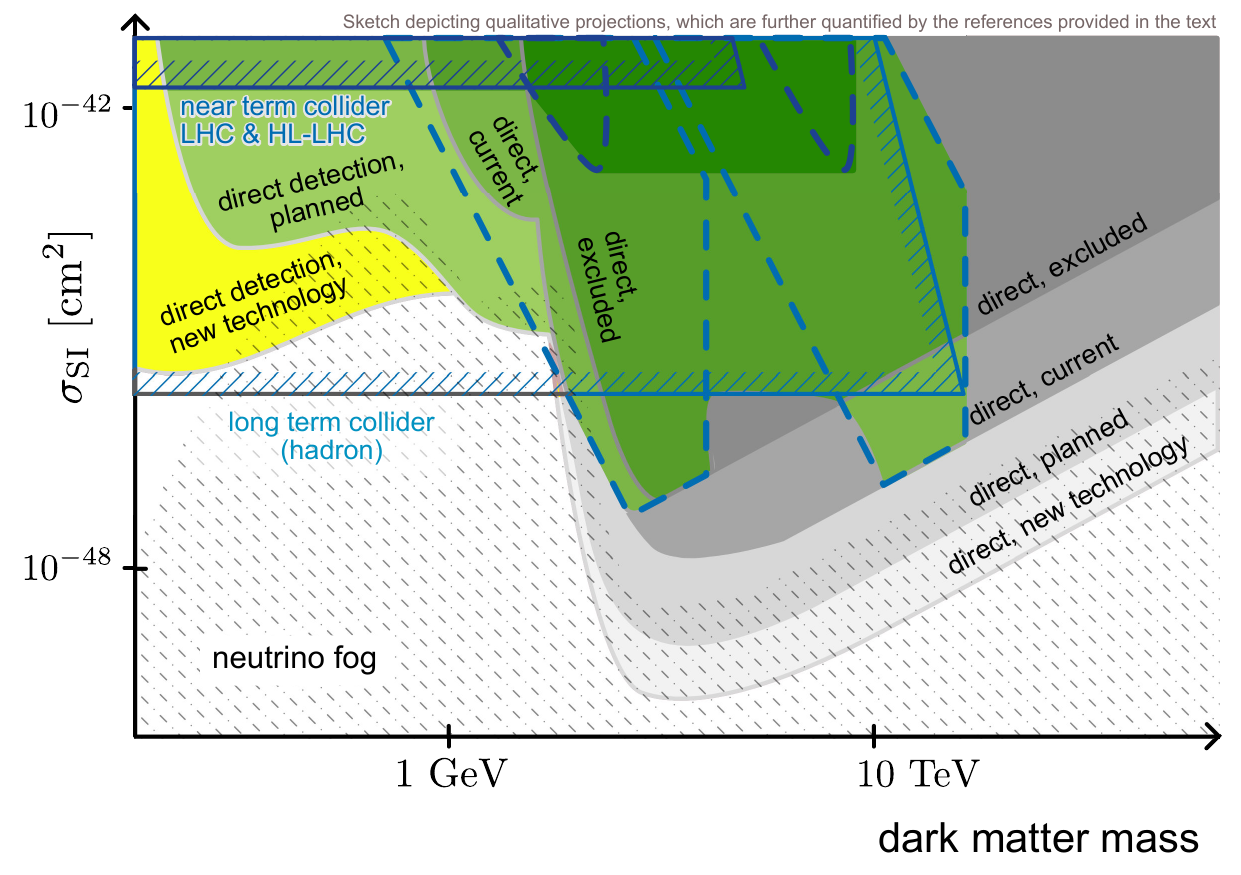}}\\
\subfloat[]{\label{subfig:BSMMediationSD}\includegraphics[width=0.7\textwidth]{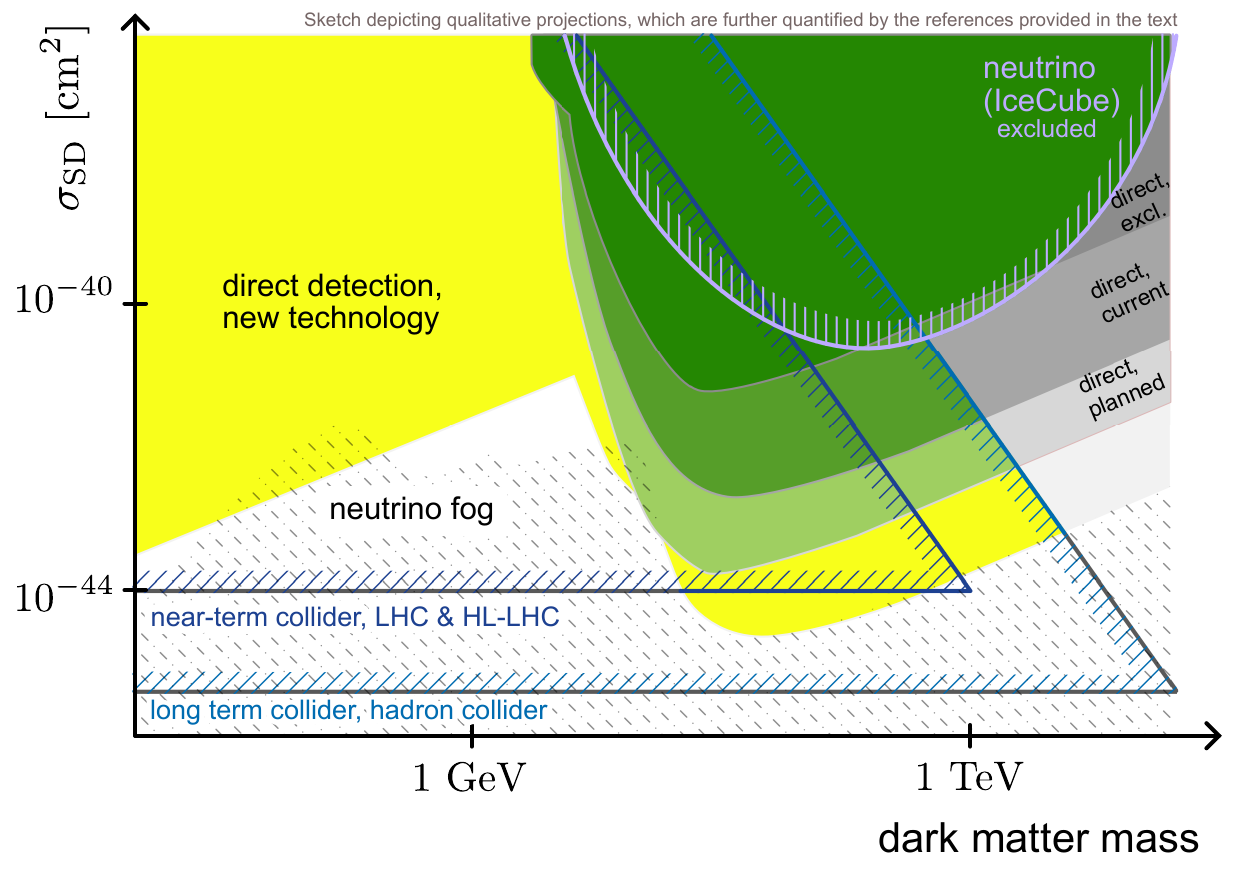}}
\end{center}
\caption{Sketches of the complementary regions of the (\protect\subref{subfig:BSMMediationSI}) spin-independent or (\protect\subref{subfig:BSMMediationSD}) spin-dependent WIMP-nucleon cross-section for a BSM-mediated simplified model of DM, as discussed in the Energy Frontier reports and contributions~\cite{Bose:2022obr,Energy-Frontier-Report,Boveia:2022jox}, where future colliders and direct detection experiments could simultaneously establish the astrophysical origin of a DM signal and study its interactions with SM particles. The qualitative coverage shown is further quantified by the references provided in the text.
The solid (fixed couplings) and dashed contours (fixed DM and/or mediator masses) illustrate the wide variance in how specific combinations of Lagrangian parameters affect the extrapolation of collider limits on the simplified model to this plane.
Also shown are regions at lower cross-section values where neutrino interactions with direct detection experiments contribute background to a search.
Regions where “excluded” is mentioned in the figure have been covered by published results, while other areas depict approximate regions of sensitivity for current and future experiments. 
Regions of overlapping coverage, where complementary observations in both types of experiments would be possible, are indicated by saturated colors. Regions accessible by one of the two types of experiments only are shown in muted colors or grayscale.
}
\label{fig:BSMMediatorDD}
\end{figure}

\begin{figure}[htp]
\begin{center}
\includegraphics[width=0.7\textwidth]{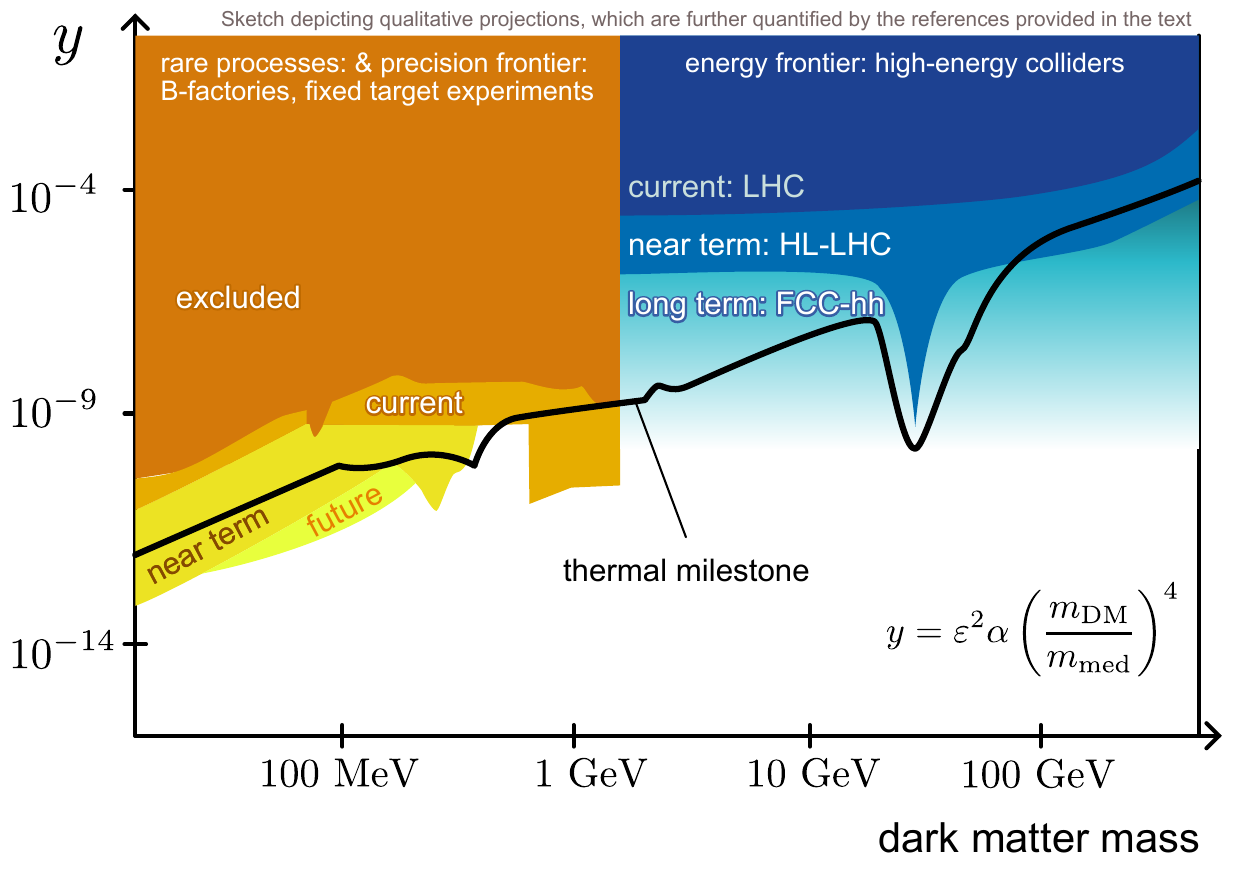}
\end{center}
\caption{Sketch of how collider and accelerator experiments together can reach sensitivity across many orders of magnitude of DM mass to couplings expected for thermal-relic vector portal inelastic Diract DM production. Shown are current and projected exclusions for both types of experiments, taken from Refs.~\cite{Bose:2022obr,Boveia:2022jox,RF6report} (note: lepton colliders not shown). The qualitative coverage shown is further quantified by the references provided in the text. The solid black line indicates the parameters which yield a thermal relic, as discussed in the frontier topical reports. 
Regions where “excluded” is mentioned in the figure have been covered by published results, while other areas depict approximate regions of sensitivity for current and future experiments. 
Regions of overlapping coverage, where complementary observations in more than one type of experiments would be possible, are indicated by saturated colors. Regions accessible by only one of the two types of experiments are shown in muted colors or grayscale.}
\label{fig:darkPhoton}
\end{figure}

Figures \ref{fig:BSMMediatorDD} and \ref{fig:darkPhoton} represent sketches of the complementarity between different DM search approaches, at collider and accelerator experiments and at Cosmic Frontier experiments. Figure \ref{fig:BSMMediatorDD} shows the complementarity in terms of simultaneous discovery, while Fig. \ref{fig:darkPhoton} shows that the unique discovery sensitivities of Energy and Rare and Precision Frontier experiments allow us to span the entire parameter space for thermal relic DM for the chosen vector portal model parameters. 
Within portal models such as the vector portal model, both low and high mass particles are needed for theoretical consistency, see e.g. \cite{Batell:2018fqo, Batell:2021xsi, Rizzo:2022qan}. 
In these cases, it is only the combination of Cosmic, Energy and Rare Processes and Precision Frontier experiment can discover both DM and the associated particle spectrum. 

\begin{figure}[htp]
\begin{center}
\includegraphics[width=0.7\textwidth]{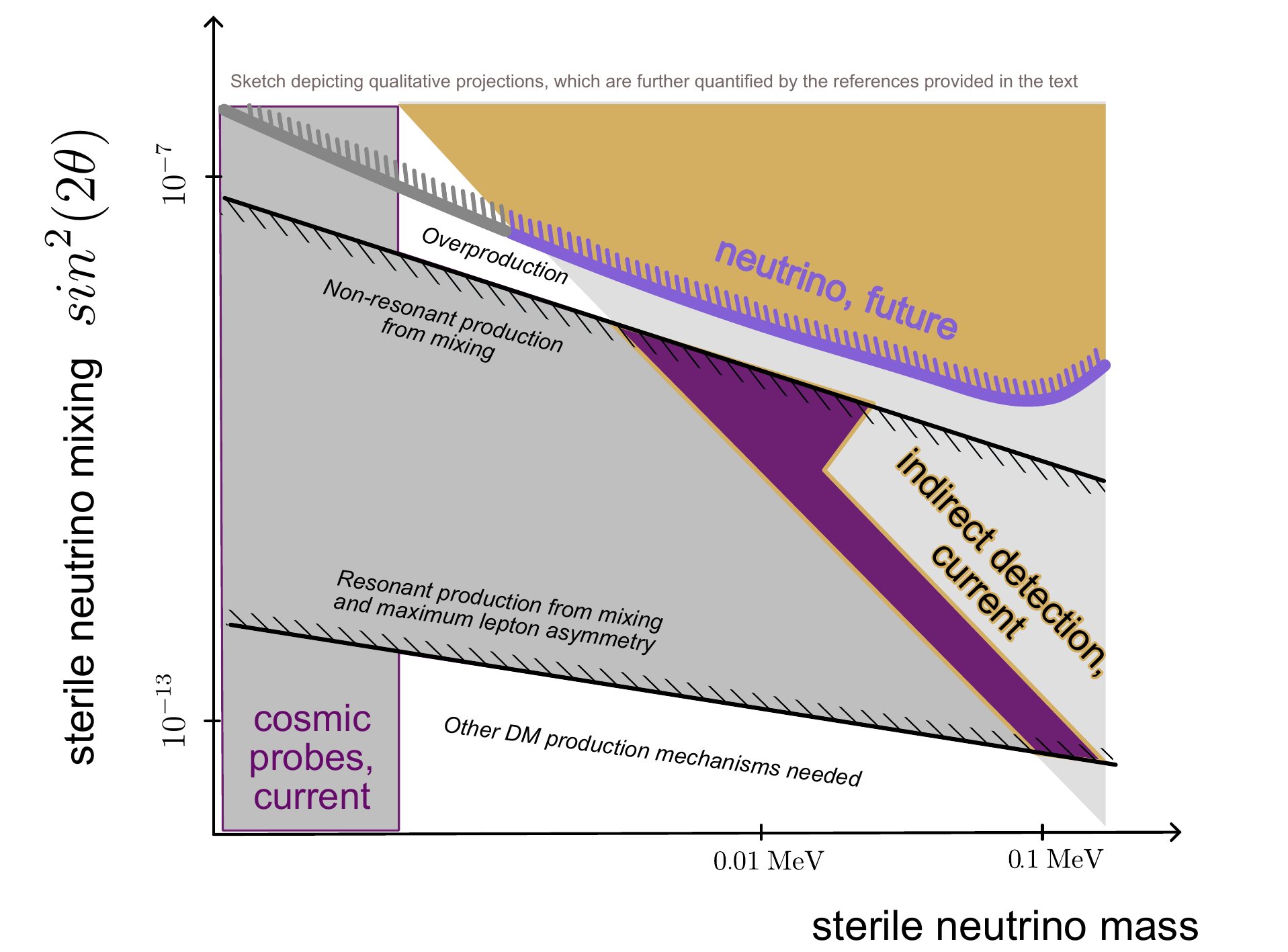}
\end{center}
\caption{Sketch of constraints on the mass and mixing angle of resonantly produced sterile neutrino dark matter from indirect detection in X-rays (e.g., \cite{Roach:2022lgo} and references therein; lighter grey and purple/orange where overlapping with other techniques), cosmic probes of small scale structure (e.g., \cite{DES:2020fxi} and references therein; darker grey and purple where overlapping with other techniques), and projected sensitivity of terrestrial tritium beta-decay neutrino experiments (orange; \cite{Merle:2017jfn} and references therein).  The qualitative coverage shown is further quantified by the references provided in the text. The solid black lines indicate the region of parameter space in which resonantly produced sterile neutrinos can constitute all of the dark matter in the neutrino minimal standard model \cite{Schneider:2016uqi,Asaka:2005an}: the upper line represents non-resonant production via active-sterile neutrino mixing, while the lower line corresponds to resonant production from mixing in the presence of the maximally allowed lepton asymmetry in the early Universe. Below this line, other production mechanisms can produce the observed amount of dark matter. 
Regions of overlapping coverage, where complementary observations in more than one type of experiments would be possible, are indicated by saturated colors. Regions accessible by only one of the two types of experiments are shown in muted colors or grayscale.
}
\label{fig:SterileNeutrino}
\end{figure}

\FloatBarrier

\subsection{Sterile Neutrino Dark Matter}
\label{sub:SterileNeutrino}


Figure~\ref{fig:SterileNeutrino} is an illustration of the complementarity between cosmological, astrophysical, and laboratory searches for sterile neutrino DM (as compiled by \cite{Abazajian:2022ofy}), with complementary reach provided by a combination of indirect DM searches (CF01; \cite{Cooley:2022ufh}), cosmic probes of structure formation (CF03; \cite{Drlica-Wagner:2022lbd}), and laboratory neutrino facilities (NF03; \cite{Coloma:2022dng}). 
Indirect detection experiments searches for X-ray lines originating from the decay of keV mass sterile neutrinos. 
Current constraints come from Chandra, XMM-Newton, NuSTAR, INTEGRAL, and Fermi GBM observations of the various astrophysical systems (e.g., the Milky Way, M31, dwarf galaxies, and galaxy clusters) \cite{Horiuchi:2013noa,Boyarsky:2005us,Boyarsky:2006zi,Abazajian:2006jc,Malyshev:2014xqa,Sicilian:2020glg,Foster:2021ngm,Roach:2019ctw}. 
Future X-ray facilities such as {\it XRISM}, {\it Athena}, and the WFM instrument aboard the {\it eXTP} X-ray Telescope could increase sensitivity to mixing angle by orders of magnitude \cite{Neronov:2015kca,Zhong:2020wre,Malyshev:2020hcc,Ando:2021fhj}.
Cosmological constraints are due to the suppression of DM structure that occurs for producing keV-mass sterile neutrinos with a ``warm'' initial momentum distribution \cite{Bullock:2017xww,Schneider:2016uqi}.
These constraints will improve by orders of magnitude as measurements of the least massive DM halos improves with DESI \cite{Valluri:2022nrh}, Rubin LSST \cite{Mao:2022fyx}, and future cosmological survey experiments \cite{Chakrabarti:2022cbu}.
Laboratory searches (e.g., Katrin, TRISTAN, BeEST, and HUNTER) \cite{Merle:2017jfn,KATRIN:2018oow,Friedrich:2020nze,Martoff:2021vxp} have different dependencies on the model behavior of sterile neutrinos in the early Universe (e.g., in cosmologies with large lepton asymmetry, low reheating temperature and/or neutrino non-standard interactions) making them highly complementary to indirect and cosmological searches \cite{Gelmini:2019clw, Gelmini:2019wfp}. Below the lowest solid line, achieving the correct relic density of sterile neutrino DM requires other production mechanisms beyond active-sterile mixing; these generally involve heavy BSM particles, which can be probed with the Energy Frontier. 

\subsection{Wave-like Dark Matter: QCD Axion}
\label{sub:axion}
Complementarity across frontiers and within the Cosmic Frontier is equally important for WLDM. However, the constraints from one frontier to another are not as stringent since there are a broader category of models to be explored both experimentally and theoretically.  For this case study, the discovery of a QCD axion by a direct detection experiment such as DMRadio-$m^3$ with a mass $< 1\,\mu$eV can be used to illustrate of the wide-ranging implication of a WLDM discovery. 

As described above, the detailed spectral measurements from direct axion searches would almost instantaneously provide a measurement of the velocity distribution of dark matter in the halo. These measurements and subsequent measurements of the position distribution can then be compared to the results from cosmic probes of dark matter.  Since axions with masses below $< 1\,\mu$eV imply additional fields at the time of inflation, CMB B-modes would then be out of range for next generation CMB experiments. A possible discrepancy between such measurements would open the door to significant changes in our understanding of particle physics and cosmology.

Most direct detection experiments use the axions' coupling to photons. A precision measurement of this coupling, or the coupling to other parts of the SM, can disentangle which category of QCD axion or ALP has been discovered. There is complementarity here between other table-top precision measurements. The discovery of a QCD axion implies additional particles, KSVZ models predict additional quarks and DFSZ predict an expanded Higgs sector, either would be strong motivation for higher-energy machines and would guide the design of such efforts.

\subsection{Selected case studies from other references}

The report on Basic Research Needs for Dark Matter Small Projects New Initiatives~\cite{BRNreport} includes three case studies for the discovery of light or ultralight DM, with example timelines and a discussion of complementarity between beam experiments, direct searches for absorption or scattering signals, and axion searches. To quote from that report, ``confirming that a detected dark-matter candidate matches the observed cosmological abundance and determining the number of dark-matter species including any subcomponents necessarily require a cross-cutting approach involving many types of experiments''. In addition to the directions discussed in Ref.~\cite{BRNreport}, future space-based MeV-band gamma-ray telescopes could potentially also have complementary sensitivity (with beam and direct-detection experiments) for MeV-scale DM candidates \cite{Coogan:2021sjs}. 
Cosmic probes of DM could identify a non-negligible DM self-interaction cross section and/or free-streaming length, both of which can naturally be large for light DM~\cite{Drlica-Wagner:2022lbd}.
Ref.~\cite{Drlica-Wagner:2022lbd} likewise includes several case studies, highlighting cosmic probes of axion-like particles that could provide a target mass for terrestrial axion searches; warm and self-interacting DM; and the prospect of detecting DM in the form of primordial black holes.

\section{Conclusions}

The nature of DM is an outstanding puzzle of fundamental physics. The space of viable, theoretically-motivated candidates is enormous and multi-dimensional, spanning many orders of magnitude in mass and interaction strength. To make progress on this challenging problem, maximize the chances of a transformative discovery, and fully elucidate the properties of DM and related new physics in the event of a discovery, we advocate a cross-Frontier effort that:
\begin{itemize}
\item incorporates multiple complementary approaches to the problem with a range of experiments at varying scales, reflecting both the sensitivity and breadth of large facilities employing mature technologies, and the potential to explore entirely new possibilities and accelerate discovery with novel approaches, 
\item includes increased dedicated support for research, including funding for collaboration that crosses project, Frontier, and/or disciplinary boundaries,
\item recognizes the essential importance of a strong and vibrant theory program -- to motivate DM searches, devise new ideas and approaches, and fully leverage cross-Frontier data to improve our understanding of DM physics -- and commits dedicated funding to support, sustain, and grow such a program. 
\end{itemize}

Complementarity between different DM searches is at the heart of this effort. The suite of approaches discussed in this summary, and in the various topical group and Frontier reports, will allow us to probe the behavior of DM across different energies and in widely-varying environments. These data will allow both sensitivity to an enormous range of diverse scenarios, and the possibility of using (both positive and null) results from multiple datasets to triangulate the nature of DM.

The past decade has seen a greater understanding of the diverse range of possibilities for DM, and simultaneously, a flourishing of new avenues for exploring its nature. The {\it next} decade will offer the opportunity to delve deep into highly compelling, long-standing, and well-studied scenarios for the nature of DM, and simultaneously to open up our search to a wide and less-explored space of exciting and well-motivated possibilities. A decade of coherent cross-frontier dark matter exploration is an opportunity that should not be missed.

\printbibliography

\section{Acknowledgments}

Research by A.~Boveia is supported by the US Department of Energy (Grant DE-SC0011726).
Research by C. Doglioni is part of projects that have received funding from the European Research Council under the European Union’s Horizon 2020 research and innovation program (grant agreement 101002463).
M. Berkat and J. Greaves's work (supervised by C. Doglioni) was produced as part of the Knut and Alice Wallenberg project Light Dark Matter (Dnr. KAW 2019.0080).
Research for P.~Harris is supported by Department of Energy Early Career Award grant DE-SC0021943.
K.Pachal's research is supported by TRIUMF, which receives federal funding via a contribution agreement with the National Research Council (NRC) of Canada.
Research by J. Shelton is supported by the US Department of Energy (Grant DE-SC0017840).
A. Steinhebel acknowledges support by an appointment to the NASA Postdoctoral Program at NASA Goddard Space Flight Center, administered by Oak Ridge Associated Universities under contract with NASA.
The work of J. Yu is supported by the U.S. Department of Energy under Grant No. DE-SC0011686.
J. Yu thanks the support of the CERN neutrino department during his stay in which majority of this work is performed.
The work of T. Slatyer is supported by the U.S. Department of Energy, Office of Science, Office of High Energy Physics under grant Contract Number DE-SC0012567, and by the National Science Foundation under Cooperative Agreement PHY-2019786 (The NSF AI Institute for Artificial Intelligence and Fundamental Interactions, http://iaifi.org/).

The authors would like to thank the following people for their input in the discussions and cross-frontier meetings that led to this report, including producing some of the summary plots: Suchita Kulkarni, Ben Loer, Bjoern Penning, Hai-Bo Yu, Jessie Shelton. 


\end{document}